\documentclass[12pt]{amsart}
\usepackage[T1]{fontenc}
\usepackage[latin1]{inputenc}
\usepackage{times}
\makeatletter

\newcommand{\LyX}{L\kern-.1667em\lower.25em\hbox{Y}\kern-.125emX\spacefactor1000}

\theoremstyle{plain}    
\newtheorem{thm}{Theorem} 
\theoremstyle{plain}    
\newtheorem{cor}{Corollary} 
\theoremstyle{plain}    
\newtheorem{lem}{Lemma} 
\theoremstyle{plain}    
\newtheorem{prop}{Proposition} 
\theoremstyle{remark}
\newtheorem{rem}{Remark}

\newcounter{const}[section]
\newcommand{\co}{c_{\thesection .\refstepcounter{const}\arabic{const}}}
\makeatother

\begin{document}

\title{On the Discrete Spectrum of a Pseudo-Relativistic Two-Body Pair Operator}

\author{Semjon Vugalter and Timo Weidl}

\address{\textsf{\tiny S. Vugalter: Mathematisches Institut der LMU, Theresienstrasse
39, 80333 Muenchen, Germany. T. Weidl: Universität Stuttgart, Fakultät Mathematik,
Pfaffenwaldring 57, D-70569 Stuttgart, Germany.}\tiny }

\date{17.11.2001}


\begin{abstract}
We prove Cwikel-Lieb-Rosenbljum and Lieb-Thirring type bounds on the discrete
spectrum of a two-body pair operator and calculate spectral asymptotics for
the eigenvalue moments and the local spectral density in the pseudo-relativistic
limit.
\end{abstract}
\maketitle

\section{Introduction}

\subsection{Statement of the problem.}

In this paper we consider the behaviour of two particles with the masses \( m_{+} \)
and \( m_{-} \) in the absence of external fields. The non-relativistic Hamiltonian
of such a system is given by
\begin{equation}
\label{schr}
-\frac{1}{2m_{+}}\Delta ^{+}-\frac{1}{2m_{-}}\Delta ^{-}-V(x^{+}-x^{-})\quad \mbox {on}\quad L^{2}(\mathbb R^{2d}),
\end{equation}
where \( x^{+},x^{-}\in \mathbb R^{d} \) denote the spatial coordinates and
\( -V \) stands for the interaction between the particles. Due to translational
invariance, this operator is unitary equivalent to the direct integral \( \int ^{\oplus }_{\mathbb R^{d}}h(P)dP \),
where
\[
h(P)=-\frac{M}{2m_{+}m_{-}}\Delta _{y}-V(y)+\frac{p^{2}}{2M},\quad p=|P|,\]
acts on \( L^{2}(\mathbb R^{d}) \). The parameter \( M=m_{+}+m_{-} \) is the
total mass of the system and \( P\in \mathbb R^{d} \) is the total momentum.
The spectrum of (\ref{schr}) is the union of the spectra of the pair operators
\( h(P) \) for all \( P\in \mathbb R^{d} \). Notice that \( h(P) \) depends
on \( P \) only by a shift of \( \frac{p^{2}}{2M} \), and the spectra of all
\( h(P) \) coincide modulo the respective shift. In other words, the fundamental
properties of the pair operator do not depend on the choice of the inertial
system of coordinates.

On the other hand, if we consider the pseudo-relativistic Hamiltonian \cite{H,LSV}
\[
\sqrt{-\Delta ^{+}+m_{+}^{2}}+\sqrt{-\Delta ^{-}+m_{-}^{2}}-V(x^{+}-x^{-}),\]
the corresponding decomposition into a direct integral \( \int ^{\oplus }_{\mathbb R^{d}}h_{rel}(P)dP \)
gives rise to the pair operators
\begin{equation}
\label{hrel}
h_{rel}(P)=\sqrt{|\mu _{+}P-i\nabla _{y}|^{2}+\mu _{+}^{2}M^{2}}+\sqrt{|\mu _{-}P+i\nabla _{y}|^{2}+\mu _{-}^{2}M^{2}}-V(y),
\end{equation}
where \( \mu _{\pm }=m_{\pm }M^{-1} \). Obviously these operators show a much
more involved dependence on the total momentum \( P\in \mathbb R^{d} \). This
implies a non-trivial behaviour of the spectra of \( h_{rel}(P) \) in \( P \).
For example, if \( -V \) is a smooth, compactly supported attractive well,
the essential spectrum of \( h_{rel}(P) \) coincides with the interval \( [(p^{2}+M^{2})^{1/2},\infty ) \)
and the discrete spectrum is finite. However, the distribution of the negative
eigenvalues of
\begin{equation}
\label{qrell}
q_{rel}(P)=h_{rel}(P)-\sqrt{p^{2}+M^{2}},\quad p=|P|,
\end{equation}
depends on \( P \). Even if the attractive force \( -V \) is too weak to induce
negative bound states for small \( p \), eigenvalues will appear as \( p \)
grows and their total number tends to infinity as \( p\to \infty  \). Our paper
is devoted to the study of this phenomenom.

More precisely, we shall study the following quantities. First, for given \( P \)
we chose the system of coordinates such that \( P=(p,0,\dots ,0) \) and we
stretch the spatial variables by the factor \( p^{-1} \). Obviously \( p^{-1}q_{rel}(P) \)
is unitary equivalent to the operator
\begin{equation}
\label{opQ}
Q(i\nabla ,y)=H_{p}(i\nabla )-V_{p}(y),
\end{equation}
where \( V_{p}(y)=V(yp^{-1}) \) and
\[
H_{p}(\xi )=T_{+}(\xi )+T_{-}(\xi )-\sqrt{1+M^{2}p^{-2}}\]
for
\[
T_{\pm }(\xi )=\sqrt{|(\eta \mp \mu _{\pm })^{2}+|\zeta |^{2}+\mu ^{2}_{\pm }M^{2}p^{-2},}\]
with \( \xi \in {\Bbb R}^{d} \), \( \xi =(\eta ,\zeta ) \) for \( \xi _{1}=\eta \in {\Bbb R} \)
and \( (\xi _{2},\dots ,\xi _{d})=\zeta \in {\Bbb R}^{d-1} \), \( \mu _{\pm }>0 \),
\( p>0 \). Throughout this paper we focus on the case of higher dimensions
\( d\geq 3 \). We will discuss the behaviour of the total number of negative
eigenvalues (including multiplicities)\footnote{
By \( \chi _{(0,\infty )} \) we denote the characteristic function of the negative
semiaxes.
}
\[
N_{p}(V)=tr\, \chi _{(-\infty ,0)}(Q_{p}(i\nabla ,y))\]
and the sum of the absolute values of the negative eigenvalues\footnote{
For real \( x \) we put \( 2x_{-}=|x|-x \).
}
\[
S_{p}(V)=tr\, (Q_{p}(i\nabla ,y))_{-}\]
of the operator \( Q_{p}(i\nabla ,y) \). In particular, we shall compare these
spectral quantities with their classical counterparts
\begin{eqnarray}
\Xi _{p}=\Xi _{p}(V) & = & (2\pi )^{-d}\int \int _{Q_{p}<0}d\xi dy,\label{phvol0} \\
\Sigma _{p}=\Sigma _{p}(V) & = & (2\pi )^{-d}\int \int (Q_{p}(\xi ,y))_{-}d\xi dy.\label{phsum0} 
\end{eqnarray}

\subsection{The classical picture.}

Already the initial analysis of the phase space averages (\ref{phvol0}) and
(\ref{phsum0}) shows somewhat unexpected results. Put \( V\geq 0 \). It is
not difficult to see, that \( \Xi _{p} \) is finite if and only if \( V\in L^{\frac{d}{2}}(\mathbb R^{d})\cap L^{d}(\mathbb R^{d}) \),
while \( \Sigma _{p} \) is finite if and only if \( V\in L^{\frac{d}{2}+1}(\mathbb R^{d})\cap L^{d+1}(\mathbb R^{d}) \).
However, within these classes of potentials the quantities \( \Xi _{p} \) and
\( \Sigma _{p} \) show various asymptotical orders in \( p \) as \( p\to \infty  \).
Indeed, we have\footnote{
Below \( \omega _{d} \) is the volume of the \( d \)-dimensional unit ball.
}
\begin{eqnarray}
\Xi _{p}(V)= & \frac{\omega _{d}p^{\frac{d+1}{2}}(1+o(1))}{2^{\frac{3d+1}{2}}\pi ^{d}}\int V^{\frac{d-1}{2}}dy & \quad \mbox {if}\quad V\in L^{\frac{d-1}{2}}\cap L^{d},\label{fo1} \\
\Sigma _{p}(V)= & \frac{\omega _{d}p^{\frac{d-1}{2}}(1+o(1))}{(d+1)2^{\frac{3d-1}{2}}\pi ^{d}}\int _{\mathbb {R}^{d}}V^{\frac{d+1}{2}}dy & \quad \mbox {if}\quad V\in L^{\frac{d+1}{2}}\cap L^{d+1}\label{fo2} 
\end{eqnarray}
as \( p\to \infty  \).\footnote{
We point out that the powers of \( V \) in (\ref{fo1}), (\ref{fo2}) are typical
for the phase space behaviour of Schrödinger operators in the spatial dimension
\( d-1 \).
} On the other hand, consider the model potentials
\begin{equation}
\label{mopo}
V_{\theta }(y)=\min \{1,v|y|^{-d/\theta }\}.
\end{equation}
If \( \frac{d-1}{2}<\theta <d \) then \( V_{\theta }\in L^{\theta }_{w}\cap L^{d}\subset (L^{\frac{d}{2}}\cap L^{d})\backslash L^{\frac{d-1}{2}} \)
and it holds
\begin{equation}
\label{fo3}
\Xi _{p}(V_{\theta })=c_{1}(d,\theta ,\mu _{\pm })p^{\theta +1}v^{\theta }M^{d-1-2\theta }(1+o(1)),\quad \frac{d-1}{2}<\theta <d,
\end{equation}
as \( p\to \infty  \), see also (\ref{tpl1}). Similarly, if \( \frac{d+1}{2}<\theta <d+1 \)
then we have \( V_{\theta }\in L^{\theta }_{w}\cap L^{d+1}\subset (L^{\frac{d}{2}+1}\cap L^{d+1})\backslash L^{\frac{d+1}{2}} \)
and
\begin{equation}
\label{fo4}
\Sigma _{p}(V_{\theta })=c_{2}(d,\theta ,\mu _{\pm })p^{\theta -1}v^{\theta }M^{d+1-2\theta }(1+o(1)),\quad \frac{d+1}{2}<\theta <d+1,
\end{equation}
as \( p\to \infty  \), cf. (\ref{tmi1}). Obviously formulae (\ref{fo3}) and
(\ref{fo4}) differ from (\ref{fo1}) and (\ref{fo2}) not only in the leading
order of \( p \), but also in the character of the dependence of the asymptotic
constants on \( V \). For the benefit of the reader we attach the calculation
of these formulae in Appendix I.

To discuss the difference in character of (\ref{fo1})-(\ref{fo2}) and (\ref{fo3})-(\ref{fo4})
it is useful to consider the massless limit case. Put
\begin{equation}
\label{qpm0}
\tilde{Q}_{p}(\xi ,y)=\tilde{H}(\xi )-V_{p}(y),
\end{equation}
where
\[
\tilde{H}(\xi )=\lim _{M\to 0}H_{p}(\xi )=|e_{+}-\xi |+|e_{-}-\xi |-2,\quad e_{\pm }=(\pm \mu _{\pm },0,\dots 0).\]
Let \( \tilde{\Xi }_{p}(V) \) and \( \tilde{\Sigma }_{p}(V) \) be the analogs
of (\ref{phvol0}) and (\ref{phsum0}), if we replace \( Q_{p} \) by \( \tilde{Q}_{p} \).
Then \( \tilde{\Xi }_{p}(V) \) and \( \tilde{\Sigma }_{p}(V) \) are finite,
if and only if \( V\in L^{\frac{d-1}{2}}\cap L^{d} \) or \( V\in L^{\frac{d+1}{2}}\cap L^{d+1} \),
respectively. For these classes of potentials the asymptotics (\ref{fo1}) and
(\ref{fo2}) can be carried over to the case \( M=0 \) as well. For potentials
(\ref{mopo}), corresponding to the cases (\ref{fo3}) or (\ref{fo4}), the
quantities \( \tilde{\Xi }_{p}(V_{\theta }) \) and \( \tilde{\Sigma }_{p}(V_{\theta }) \)
are infinite for all \( p>0 \).

\subsection{Estimates on the counting function. }

In section 3 we start the spectral analysis of the operators (\ref{opQ}) and
develop Cwikel-Lieb-Rosenbljum type bounds on the counting function \( N_{p}(V) \).
The strong inhomogeneity of the symbol prevents us from using ready standard
versions of Cwikel inequality \cite{C,BKS}. Instead we apply a modification
\cite{W1,W2}, where the estimate follows the phase space distribution as close
as possible even for complicated symbols. In particular, we show that for \( p\geq M>0 \)
\begin{eqnarray}
N_{p}(V) & \leq  & c\left( p^{2}\left( 1+\ln pM^{-1}\right) \left\Vert V\right\Vert _{L^{1}}+\left\Vert V\right\Vert _{L^{3}}^{3}\right) ,\quad d=3,\label{esnp1} \\
N_{p}(V) & \leq  & c\left( p^{\frac{d+1}{2}}\left\Vert V\right\Vert _{L^{\frac{d-1}{2}}}^{\frac{d-1}{2}}+\left\Vert V\right\Vert _{L^{d}}^{d}\right) ,\quad d\geq 4,\label{esnp2} \\
N_{p}(V) & \leq  & c\left( p^{1+\theta }M^{d-1-2\theta }\left\Vert V\right\Vert ^{\theta }_{\theta ,w}+\left\Vert V\right\Vert _{L^{d}}^{d}\right) ,\quad d\geq 3,\label{esnp3} 
\end{eqnarray}
where \( \frac{d-1}{2}<\theta <d \) in (\ref{esnp3})\footnote{
Here \( \left\Vert \cdot \right\Vert _{\theta ,w} \) stands for the ``weak''
norm of the Lorentz space \( L^{\theta }_{w} \).
}, whenever the respective r.h.s. is finite. The leading terms in the bounds
(\ref{esnp2}) and (\ref{esnp3}) reduplicate the correct asymptotic order in
\( p \) in (\ref{fo1}) and (\ref{fo3}). 

The appearance of some mass dependence in (\ref{esnp1}) is natural, since one
expects that the massless operator \( \tilde{Q}_{p} \) has generically infinite
negative spectrum for \( d=3 \) and all \( p>0 \). Indeed, the massless kinetic
energy \( \tilde{H}(\xi ) \) vanishes on the interval between \( e_{+} \)
and \( e_{-} \), the first coordinate of the momentum will not contribute in
this region and we experience practically a \( d-1 \) dimensional kinetic behaviour.
Hence, to establish (\ref{esnp1}) for \( d=3 \) we have to deal with problems
resembling spectral estimates for two-dimensional Schrödinger operators. In
the massless case virtual bound states will prevent any estimates on \( N_{p}(V) \).
The inclusion of a finite mass supresses this effect to some extend, but leads
with our method of proof to the additional factor \( (1+\ln pM^{-1}) \) in
(\ref{esnp1}) compared to (\ref{fo1}). 

If the potential \( V \) has a repulsive tail at infinity, the bound (\ref{esnp1})
can be complemented by the estimate
\[
N_{p}\leq c(V)p^{2},\quad p\geq M>0,\quad d=3.\]
This is carried out in Theorem \ref{tm: A1} in Appendix II. Moreover, combining
the techniques of Appendix II and inequality (\ref{esnp1}) it is possible to
show that \( N=o(p^{2}\ln pM^{-1}) \) as \( p\to \infty  \) for arbitrary
\( V\in L^{1}(\mathbb R^{3})\cap L^{3}(\mathbb R^{3}) \). Nevertheless it remains
an open problem, up to what extend the logarithmic increase in \( p \) can
be removed from (\ref{esnp1}) in general.

\subsection{Estimates on the eigenvalue moments.}

In section 4 we integrate the estimates (\ref{esnp1})-(\ref{esnp3}) according
to the Lieb-Aizenman trick \cite{AL} to obtain Lieb-Thirring type bounds on
the sums of the negative eigenvalues and find that for \( p\geq M>0 \)
\begin{eqnarray}
S_{p}(V) & \leq  & c\left( p\left( 1+\ln pM^{-1}\right) \left\Vert V\right\Vert _{L^{2}}+p^{-1}\left\Vert V\right\Vert _{L^{4}}^{4}\right) ,\quad d=3,\label{essp1} \\
S_{p}(V) & \leq  & c\left( p^{\frac{d-1}{2}}\left\Vert V\right\Vert _{L^{\frac{d+1}{2}}}^{\frac{d+1}{2}}+p^{-1}\left\Vert V\right\Vert _{L^{d+1}}^{d+1}\right) ,\quad d\geq 4,\label{essp2} \\
S_{p}(V) & \leq  & c\left( p^{1-\theta }M^{d+1-2\theta }\left\Vert V\right\Vert ^{\theta }_{\theta ,w}+p^{-1}\left\Vert V\right\Vert _{L^{d+1}}^{d+1}\right) ,\quad d\geq 3,\label{essp3} 
\end{eqnarray}
where \( \frac{d+1}{2}<\theta <d \) in (\ref{essp3}). The bounds (\ref{essp1})
and (\ref{essp2}) are immediate consequences of (\ref{esnp1}) and (\ref{esnp2}),
respectively. The estimate (\ref{essp1}) carries again an additional logarithmic
factor. Since eigenvalue moments behave usually more regular than counting functions,
the question on the essence of this term stands even more pressing in this situation.
The derivation of (\ref{essp3}) from (\ref{esnp3}) is somehow more involved,
because bounds with Lorentz norms cannot be handled in the same way as in \cite{AL}.

\subsection{Spectral asymptotics and coherent states.}

In section 5 we state in Theorems \ref{sec: maintm1} and \ref{sec: maintm2}
the main asymptotic results of this paper. In a first step we obtain the formula
\begin{equation}
\label{asspmain}
S_{p}(V)=(1+o(1))\Sigma _{p}(V)\quad \mbox {as}\quad p\to \infty ,
\end{equation}
if for \( d=3 \) the potential \( V \) has uniformly bounded, continuous second
derivatives and \( V\in L^{\theta }(\mathbb {R}^{3})\cap L^{4}(\mathbb {R}^{3}) \)
for some \( \theta <2 \); or if \( V\in L^{\frac{d+1}{2}}(\mathbb {R}^{d})\cap L^{d+1}(\mathbb {R}^{d}) \)
for \( d\geq 4 \). This result, which is obtained by means of coherent states,
corresponds essentially to the case of the phase space asymptotics (\ref{fo2})
and relates to the bounds (\ref{essp1}), (\ref{essp2}). In section 5 we provide
the necessary background information on Berezin-Lieb inequalities. In sections
6 and 7 we implement these methods for the specific symbol at hand. The proof
of Theorem \ref{sec: maintm1} is finally given in section 8. We point out that
our methods do not avail for spectral asymptotics in the case (\ref{fo4}).

While the coherent state method works well for traces of convex functions of
the operator, such as \( S_{p}(V) \), the application to counting functions
is more subtle. Essentially one has to differentiate the asymptotic formula
(\ref{asspmain}), what requires special attention. In section 9 we avail to
the extend, that we can give asymptotics of the local spectral density. Assume
that \( U,V\geq 0 \), \( U,V\in L^{\theta }\cap L^{d+1} \) for some \( \theta <\frac{d+1}{2} \)
and that \( U \) and \( V \) possess uniformly bounded second derivatives.
Put \( U(y;p)=U(p^{-1}y) \). Then
\[
\lim _{p\to \infty }p^{-\frac{d+1}{2}}tr\: U(y;p)\chi _{0}(Q_{p}(i\nabla ,y))=\frac{\omega _{d}}{2^{\frac{3d+1}{2}}\pi ^{d}}\int U(x)V^{\frac{d-1}{2}}(x)dx.\]
The function \( U \) has to decay at infinity and one cannot put \( U=1 \)
and deduce an asymptotic for \( N_{p}(V) \) itself. However, it is clear that
\[
\liminf _{p\to \infty }p^{-\frac{d+1}{2}}N_{p}(V)\geq \frac{\omega _{d}}{2^{\frac{3d+1}{2}}\pi ^{d}}\int V^{\frac{d-1}{2}}dx.\]
This sharp lower bound complements the estimates from above (\ref{esnp1}) and
(\ref{esnp2}). We follow an approach similar to \cite{ELSS}. Our methods do
not provide sharp asymptotics in the setting of (\ref{fo3}).

\subsection{Acknowledgements. }

The authors acknowledge the financial support from the Swedish Institute, the
DAAD and the EU Network on Quantum Mechanics. We wish also to express our gratidute
to H. Siedentop for the numerous fruitful discussions on this material. The
first author acknowledges the kind support of ESI Vienna.

\section{Notation}

Let \( L^{p}(\mathbb {R}^{d}) \) be the space of \( p \)-integrable functions
with respect to the Lebesgue measure \( \nu =dx \) on \( \mathbb {R}^{d} \)
equipped with the standard norm \( \left\Vert \cdot \right\Vert _{L^{p}(\mathbb {R}^{d})} \).
We shall omit the spaces from our notation where possible.

If \( f \) is a real-valued function on \( \mathbb {R}^{d} \) and measurable
with respect to the Lebesgue measure \( \nu  \), then put
\begin{eqnarray}
f_{\pm }(x) & = & (|f(x)|\pm f(x))/2,\\
\nu _{f}(s) & = & \nu \left( \{|f(x)|>s\}\right) ,\quad s>0,\label{nuf} \\
f^{*}(t) & = & \inf _{\nu _{f}(s)\leq t}s,\quad t>0.\label{rearr} 
\end{eqnarray}
Note that \( \int |f|^{q}d\nu =\int (f^{*})^{q}dt \) and that \( |f_{1}(x)|\geq |f_{2}(x)| \)
for a.e. \( x\in \mathbb {R}^{d} \) implies \( f_{1}^{*}(t)\geq f_{2}^{*}(t) \)
for all \( t>0 \). We say that \( f\in L^{q}_{w}(\mathbb {R}^{d}) \) if
\[
\left\Vert f\right\Vert _{q,w}=\sup _{t>0}t^{-q^{-1}}f^{*}(t)\]
is finite. Beside the quasi-norm \( \left\Vert \cdot \right\Vert _{q,w} \)
we shall also use the asymptotical functionals
\begin{eqnarray*}
\delta _{q}(f) & = & \liminf _{t\to \infty }t^{-q^{-1}}f_{\nu }^{*}(t),\\
\Delta _{q}(f) & = & \limsup _{t\to \infty }t^{-q^{-1}}f_{\nu }^{*}(t),
\end{eqnarray*}
which are continuous on \( L^{q}_{w}(\mathbb {R}^{d}) \).

The function \( \chi _{M} \) will denote the characteristic function of the
set \( M \). If \( M=(-\infty ,t)\subset \mathbb {R} \) we write in shorthand
\( \chi _{t}=\chi _{(-\infty ,t)} \). Let \( \omega _{d} \) stand for the
volume of the unit ball in \( \mathbb R^{d} \).

Finally, by \( c \) or \( c_{j.k} \) we denote various constants where we
do not keep track of their exact values. In particular, the same notion \( c \)
in different equations does not imply that these constants coincide.

\section{Uniform Estimates on the Number of Negative Eigenvalues: Cwikel's Inequality
Revised}

\subsection{Statement of the result.}

\label{sec: 3}In this section we discuss a priori bounds on the counting function
of the discrete spectrum of the operator
\[
Q_{p}(i\nabla ,y)=H_{p}(i\nabla )-V_{p}(y).\]
Our goal is to find estimates, which reproduce the behaviour of the phase space
\[
\Xi _{p}=\Xi _{p}(V)=(2\pi )^{-d}\int \int _{Q_{p}<0}d\xi dy\]
in general, and the asymptotics of \( \Xi _{p} \) for \( p\to \infty  \) in
particular, as closely as possible. In particular, we shall obtain the following
two statements.

\begin{thm}
\label{sec: tm1}Assume that \( V\geq 0 \), \( V\in L^{\frac{d-1}{2}}(\mathbb {R}^{d})\cap L^{d}(\mathbb {R}^{d}) \)
and \( p\geq M>0 \). Then there exists a finite constant \( c=c(d) \), which
is independent on \( p \), \( M \) and \( V \), such that
\begin{eqnarray}
N_{p}(V) & \leq  & c\left( p^{2}\left( 1+\ln pM^{-1}\right) \left\Vert V\right\Vert _{L^{1}}+\left\Vert V\right\Vert _{L^{3}}^{3}\right) ,\quad d=3,\label{bd13} \\
N_{p}(V) & \leq  & c\left( p^{\frac{d+1}{2}}\left\Vert V\right\Vert _{L^{\frac{d-1}{2}}}^{\frac{d-1}{2}}+\left\Vert V\right\Vert _{L^{d}}^{d}\right) ,\quad d\geq 4.\label{bd13d4} 
\end{eqnarray}

\end{thm}
\begin{rem}
Note that for \( d=3 \) in contrast to the asymptotical behaviour of the phase
space volume \( \Xi _{p}\asymp p^{2}\left\Vert V\right\Vert _{L^{1}} \) as
\( p\to \infty  \) for \( V\in L^{1}(\mathbb R^{3})\cap L^{3}(\mathbb R^{3}) \),
the bound (\ref{bd13}) contains an additional logarithmic factor. This underlines,
that formula (\ref{bd13}) has in fact a two-dimensional character, see {[}W2{]}. 
\end{rem}
\vspace{-0mm}

\begin{rem}
We point out, that in the case \( M=0 \) in the dimension \( d=3 \) one expects
infinite many negative eigenvalues for any non-trivial attractive potential
\( V\geq 0 \). In contrast to that in higher dimensions the bound (\ref{bd13d4})
holds true in the massless case as well.
\end{rem}
\begin{thm}
\label{sec: tm2}Assume that \( d\geq 3 \), \( V\geq 0 \) and \( V\in L^{\theta }_{w}(\mathbb {R}^{d})\cap L^{d}(\mathbb {R}^{d}) \)
for \( \frac{d-1}{2}<\theta <\frac{d}{2} \). Then there exist finite constants
\( c_{1}(\theta ) \) and \( c_{2}(\theta ) \) independent on \( p \), \( M \)
and \( V \), such that
\begin{equation}
\label{the11}
N_{p}(V)\leq c_{1}(\theta )p^{1+\theta }M^{d-1-2\theta }\left\Vert V\right\Vert ^{\theta }_{\theta ,w}+c_{2}(\theta )\left\Vert V\right\Vert _{L^{d}}^{d}
\end{equation}
for all \( 0<M\leq p \).
\end{thm}
\begin{rem}
The corresponding asymptotics shows that for large \( p \) the r.h.s. of (\ref{the11})
is of the same order in \( p \) as \( \Xi _{p}(V) \), if the potential \( V \)
satisfies \( \delta _{\theta }(V)=\Delta _{\theta }(V)=v>0 \).
\end{rem}
The remaining part of this section is devoted to the proof of Theorem \ref{sec: tm1}
and Theorem \ref{sec: tm2}.

\subsection{A modification of Cwikel's inequality}

Let \( Q_{A,B} \) be an operator of the type
\[
Q_{A,B}=B(i\nabla )-A(x)\]
on \( L^{2}(\mathbb {R}^{d}) \), where \( A=a^{2} \) and \( B=b^{-2} \) with
\( a,b\geq 0 \). Assume that the operator
\[
E_{a,b}=a(x)b(i\nabla )\]
is compact in \( L^{2}(\mathbb {R}^{d}) \) and let \( \{s_{n}(E_{a,b})\}_{n\geq 1} \)
be the non-increasing sequence of the singular values (approximation numbers)
of \( E_{a,b} \). According to the Birman-Schwinger principle \cite{B,S} the
total multiplicity of the negative spectrum of \( Q_{A,B} \) equals to the
number of singular values \( s_{n}(E_{a,b}) \) exceeding one, that is
\[
N_{A,B}:=tr\, \chi _{0}(Q_{A,B})=\mbox {card}\, \left\{ n:s_{n}(E_{a,b})>1\right\} .\]
Hence, spectral estimates on the operators \( Q_{A,B} \) can be found in terms
of estimates on the sequence \( \{s_{n}(E_{a,b})\}_{n\geq 1} \). In particular,
if \( a \) and \( b \) satisfy \( a\in L^{r}(\mathbb {R}^{d}) \) and \( b\in L^{r}_{w}(\mathbb {R}^{d}) \)
for some \( 2<r<\infty  \), then according to {[}C{]} \( E_{a,b}\in S_{\infty }(L^{2}(\mathbb {R}^{d})) \)
and
\begin{equation}
\label{CLR}
s_{n}(E_{a,b})\leq \co (r,d)n^{-1/r}\left\Vert a\right\Vert _{L^{r}}\left\Vert b\right\Vert _{r,w}\quad \mbox {for}\: \mbox {all}\quad n\in \mathbb {N}.
\end{equation}

The bound (\ref{CLR}) is of particular interest if \( b(\xi )=|\xi |^{-d/r}\in L^{r}_{w}(\mathbb {R}^{d}) \),
since then the factor \( \left\Vert a\right\Vert _{L^{r}}^{r} \) is proportional
to the volume of the portion of the classical phase space given by
\[
\{(x,\xi )\in \mathbb {R}^{d}\times \mathbb {R}^{d}|a(x)b(\xi )>1\}.\]
For functions \( b(\xi ) \) which are not ``optimal'' members of the weak
class \( L^{r}_{w}(\mathbb {R}^{d}) \), the right hand side of (\ref{CLR})
does not capture the respective phase space volumina. We are therefore in need
for a suitable generalisation of (\ref{CLR}), which is applicable to a sufficiently
wide class of symbols \( b \) and which reflects the phase space character
of the estimate even for non-homogeneous symbols. Corresponding results can
be found in \cite{W1,W2}. For the problem at hand we shall use the following
statement from \cite{W2}.

Consider the function \( q(x,\xi )=a(x)b(\xi ) \) on \( \mathbb {R}^{d}\times \mathbb {R}^{d} \)
and assume that \( q\in L^{2}(\mathbb {R}^{2d})+L_{0}^{\infty }(\mathbb {R}^{2d}) \).
Here \( L_{0}^{\infty }(\mathbb {R}^{2d}) \) stands for the subspace of bounded
functions \( q \) satisfying \( q(x,\xi )\to 0 \) as \( |x|+|\xi |\to \infty  \).
Let \( q^{*} \) be the non-increasing rearrangement of \( q \), see (\ref{rearr})
and put
\begin{equation}
\label{qqq}
\left\langle q\right\rangle (\hat{t})=\left( \hat{t}^{-1}\int _{0}^{\hat{t}}(q^{*}(t))^{2}dt\right) ^{1/2},
\end{equation}
which is finite for any \( \hat{t}>0 \). If \( \nu =dxd\xi  \) is the Lebesgue
measure on \( \mathbb {R}^{2d} \) and the distribution function \( \nu _{q} \)
is defined according to (\ref{nuf}), then using integration by parts the quantity
(\ref{qqq}) can also be rewritten as follows
\begin{equation}
\label{qq}
\left\langle q\right\rangle (\hat{t})=\left( (q^{*}(\hat{t}))^{2}+\frac{2}{\hat{t}}\int _{q^{*}(\hat{t})}^{\infty }s\nu _{q}(s)ds\right) ^{1/2},\quad \hat{t}>0.
\end{equation}
The following proposition holds true:

\begin{prop}
\label{pro1}(\cite{W2}) Assume that \( q(x,\xi )=a(x)b(\xi )\in L^{2}(\mathbb {R}^{2d})+L_{0}^{\infty }(\mathbb {R}^{2d}) \).
Then \( E_{a,b}\in S_{\infty }(L^{2}(\mathbb {R}^{d})) \) and the inequality
\begin{equation}
\label{TWCLR}
s_{n}(E_{a,b})\leq 5\left\langle q\right\rangle ((2\pi )^{d}n)
\end{equation}
holds true for all \( n\in \mathbb {N} \). 
\end{prop}
\begin{rem}
In conjunction with the Birman-Schwinger principle the bound (\ref{TWCLR})
implies
\begin{equation}
\label{BS}
\frac{1}{5}\leq \left\langle q\right\rangle \left( (2\pi )^{d}N_{A,B}\right) .
\end{equation}

\end{rem}

\subsection{Cwikel´s inequality for the operator \protect\( H_{p}(\xi )-V_{p}(y)\protect \).
Preliminary estimates.}

Now we apply Proposition \ref{pro1} to the particular symbol \( q_{p}(x,\xi )=a_{p}(x)b_{p}(\xi ) \)
with \( A_{p}(x)=a_{p}^{2}(x)=V_{p}(x)\geq 0 \) and \( B_{p}(\xi )=b_{p}^{-2}(\xi )=H_{p}(\xi ) \).
We start with some basic observations. Obviously it holds
\[
\nu _{q_{p}}(s)=\nu \left\{ (x,\xi )\in \mathbb {R}^{d}\times \mathbb {R}^{d}|q_{p}(x,\xi )>s\right\} =\Xi _{p}(s^{-2}V),\quad s>0.\]
The behaviour of the quantity \( \Xi _{p} \) is analysed in Appendix I. We
establish there that according to (\ref{prelest}) and (\ref{Ladef1}) for \( p\geq M \)
the two-sided bound
\begin{equation}
\label{nunu}
\nu _{q_{p}}(s)\asymp \nu _{q_{p,1}}(s)+\nu _{q_{p,2}}(s)+\nu _{q_{p,3}}(s)
\end{equation}
holds true, where
\begin{eqnarray}
\nu _{q_{p,1}}(s) & = & \frac{p^{\frac{d}{2}+1}}{s^{d}M}\int _{\Omega _{1}(p,s)}V^{\frac{d}{2}}dx\label{nunu1} \\
\nu _{q_{p,2}}(s) & = & \frac{p^{\frac{d+1}{2}}}{s^{d-1}}\int _{\Omega _{2}(p,s)}V^{\frac{d-1}{2}}dx,\label{nunu2} \\
\nu _{q_{p,3}}(s) & = & s^{-2d}\int _{\Omega _{3}(p,s)}V^{d}dx,\label{nunu3} 
\end{eqnarray}
and
\begin{eqnarray}
\Omega _{1}(p,s) & = & \{x|V(x)\leq s^{2}M^{2}p^{-1}\},\label{O1} \\
\Omega _{2}(p,s) & = & \{x|s^{2}M^{2}p^{-1}<V(x)\leq s^{2}p\},\label{O2} \\
\Omega _{3}(p,s) & = & \{x|V(x)>s^{2}p\}.\label{O3} 
\end{eqnarray}

Moreover, note that from (\ref{nunu}) and (\ref{O1}), (\ref{O2}) one concludes
\[
\nu _{q_{p}}(s)\geq \co \frac{p^{d+1}}{s^{2d}M^{d+1}}\int _{\Omega _{1}(p,s)}V^{d}dx+\co s^{-2d}\int _{\Omega _{2}(p,s)\cup \Omega _{3}(p,s)}V^{d}dx,\quad s>0.\]
Since we assume \( p\geq M \), the bound \( \nu _{q_{p}}(s)\geq \co s^{-2d}\left\Vert V\right\Vert ^{d}_{L^{d}} \)
holds true. Hence, for the inverse \( q^{*}_{p} \) of \( \nu _{q_{p}} \) we
have
\begin{equation}
\label{O4}
q_{p}^{*}(t)\geq \co t^{-\frac{1}{2d}}\left\Vert V\right\Vert _{L^{d}}^{1/2},\quad t>0.
\end{equation}

\subsection{Potentials \protect\( V\in L^{\frac{d-1}{2}}(\mathbb {R}^{d})\cap L^{d}(\mathbb {R}^{d})\protect \).}

For this class of potentials (\ref{nunu}) and (\ref{O1}) imply
\[
\nu _{q_{p}}(s)\leq \co \max \left\{ s^{1-d}p^{\frac{d+1}{2}}\left\Vert V\right\Vert ^{\frac{d-1}{2}}_{L^{\frac{d-1}{2}}},s^{-2d}\left\Vert V\right\Vert ^{d}_{L^{d}}\right\} ,\]
or
\begin{equation}
\label{O5}
q_{p}^{*}(t)\leq \co \max \left\{ t^{-\frac{1}{d-1}}p^{\frac{d+1}{2(d-1)}}\left\Vert V\right\Vert ^{1/2}_{L^{\frac{d-1}{2}}},t^{-\frac{1}{2d}}\left\Vert V\right\Vert ^{1/2}_{L^{d}}\right\} .
\end{equation}
Assume now that \( d\geq 4. \) Then (\ref{O5}), (\ref{qqq}) and (\ref{BS})
imply
\begin{equation}
\label{O6}
1\leq \co (N_{p}(V))^{-\frac{2}{d-1}}p^{\frac{d+1}{d-1}}\left\Vert V\right\Vert _{L^{\frac{d-1}{2}}}+\co (N_{p}(V))^{-\frac{1}{d}}\left\Vert V\right\Vert _{L^{d}}.
\end{equation}

The analogous bound for the case \( d=3 \) requires some more attention. For
this we insert each of the three summand (\ref{nunu1})-(\ref{nunu3}) in (\ref{nunu})
into the integral in (\ref{qq}) and obtain
\begin{eqnarray}
\frac{1}{\hat{t}}\int ^{\infty }_{q_{p}^{*}(\hat{t})}s\nu _{1,q_{p}}(s)ds & \leq  & \co \hat{t}^{-1}M^{-1}p^{\frac{5}{2}}\int _{\mathbb {R}^{3}}dxV^{\frac{3}{2}}(x)\int _{M^{-1}\sqrt{pV(x)}}^{\infty }s^{-2}ds\notag \\
 & \leq  & \co \hat{t}^{-1}p^{2}\left\Vert V\right\Vert _{L^{1}},\label{p1} 
\end{eqnarray}

\begin{eqnarray}
\frac{1}{\hat{t}}\int ^{\infty }_{q_{p}^{*}(\hat{t})}s\nu _{2,q_{p}}(s)ds & \leq  & \co \hat{t}^{-1}p^{2}\int _{\mathbb {R}^{3}}dxV(x)\int _{\sqrt{p^{-1}V(x)}}^{M^{-1}\sqrt{pV(x)}}s^{-1}ds\notag \\
 & \leq  & \co \left\Vert V\right\Vert _{L^{1}}\hat{t}^{-1}p^{2}\ln pM^{-1},\label{p2} 
\end{eqnarray}
as well as
\begin{eqnarray*}
\frac{1}{\hat{t}}\int ^{\infty }_{q_{p}^{*}(\hat{t})}s\nu _{3,q_{p}}(s)ds & \leq  & \co \hat{t}^{-1}\int _{\mathbb {R}^{3}}dxV^{3}(x)\int _{q^{*}_{p}(\hat{t})}^{\infty }s^{-5}ds\\
 & \leq  & \co \hat{t}^{-1}(q_{p}^{*}(\tau ))^{-4}\left\Vert V\right\Vert _{L^{3}}^{3}.
\end{eqnarray*}
By (\ref{O4}) the last bound implies
\begin{equation}
\label{p3}
\frac{1}{\hat{t}}\int ^{\infty }_{q_{p}^{*}(\hat{t})}s\nu _{3,q_{p}}(s)ds\leq \co \hat{t}^{-1/3}\left\Vert V\right\Vert _{L^{3}}.
\end{equation}
If we insert (\ref{O5})-(\ref{p3}) into (\ref{qq}) and (\ref{BS}) we arrive
at
\begin{equation}
\label{bd130}
1\leq \co \max \left\{ (N_{p}(V))^{-1}\left\Vert V\right\Vert _{L^{1}}p^{2}\left( 1+\ln pM^{-1}\right) ,(N_{p}(V))^{-1/3}\left\Vert V\right\Vert _{L^{3}}\right\} .
\end{equation}
The relations (\ref{O6}) and (\ref{bd130}) imply Theorem \ref{sec: tm1}.

\subsection{Potentials \protect\( V\in L^{\theta }_{w}(\mathbb R^{d})\cap L^{d}(\mathbb R^{d})\protect \),
\protect\( \frac{d-1}{2}<\theta <\frac{d}{2}\protect \).}

First observe, that (\ref{nunu3}) implies
\begin{equation}
\label{v3co}
\nu _{3,q_{p}}(s)\leq \co s^{-2d}\left\Vert V\right\Vert _{L^{d}}^{d},\quad s>0.
\end{equation}
Furthermore, by (\ref{nunu1}) and (\ref{nunu2}) we have
\begin{eqnarray}
 &  & \notag \nu _{1,q_{p}}(s)+\nu _{2,q_{p}}(s)\\
 & \leq  & \co \int _{\mathbb R^{d}}\min \left\{ \frac{p^{\frac{d}{2}+1}}{s^{d}M}V^{\frac{d}{2}}(x),\frac{p^{\frac{d+1}{2}}}{s^{d-1}}V^{\frac{d-1}{2}}(x)\right\} dx.\label{OO} 
\end{eqnarray}
Assume now \( \left\Vert V\right\Vert _{L_{w}^{\theta }}\leq v \), that is
\( V^{*}(t)\leq vt^{-\frac{1}{\theta }} \) for all \( t>0 \). Passing from
integration in space to integration of rearrangements (\ref{OO}) turns into
\begin{eqnarray*}
\nu _{1,q_{p}}(s)+\nu _{2,q_{p}}(s) & \leq  & \co \int _{0}^{\infty }\min \left\{ \frac{p^{\frac{d}{2}+1}}{s^{d}M}v^{\frac{d}{2}}t^{-\frac{d}{2\theta }},\frac{p^{\frac{d+1}{2}}}{s^{d-1}}v^{\frac{d-1}{2}}t^{-\frac{d-1}{2\theta }}\right\} dt\\
 & \leq  & \co (\theta )\frac{p^{\frac{d}{2}+1}}{s^{d}M}v^{\frac{d}{2}}t_{c}^{1-\frac{d}{2\theta }}+\co (\theta )\frac{p^{\frac{d+1}{2}}v^{\frac{d-1}{2}}}{s^{d-1}}t_{c}^{1-\frac{d-1}{2\theta }}
\end{eqnarray*}
with \( t_{c}=M^{-2\theta }s^{-2\theta }v^{\theta }p^{\theta } \), and
\[
\nu _{1,q_{p}}(s)+\nu _{2,q_{p}}(s)\leq \co (\theta )v^{\theta }M^{d-1-2\theta }s^{-2\theta }p^{1+\theta }.\]
Together with (\ref{v3co}) this gives
\[
\nu _{q_{p}}(s)\leq \co (\theta )\max \left\{ p^{1+\theta }M^{d-1-2\theta }s^{-2\theta }\left\Vert V\right\Vert _{\theta ,w}^{\theta },s^{-2d}\left\Vert V\right\Vert _{L^{d}}^{d}\right\} ,\quad s>0,\]
and
\[
q^{*}_{p}(t)\leq \co (\theta )\max \left\{ p^{\frac{1}{2}+\frac{1}{2\theta }}M^{\frac{d-1}{2\theta }-1}\left\Vert V\right\Vert _{\theta ,w}^{\frac{1}{2}}t^{-\frac{1}{2\theta }},\left\Vert V\right\Vert _{L^{d}}^{\frac{1}{2}}t^{-\frac{1}{2d}}\right\} ,\quad t>0.\]
From (\ref{BS}) we conclude Theorem \ref{sec: tm2}.

\section{Uniform estimates on the Eigenvalue Moments: Lieb-Thirring Inequalities Revised.}

\subsection{Statement of the results.}

\label{sec: 4}Alongside with estimates on the number of negative eigenvalues
we shall make use of estimates on the moments of eigenvalues. Given a bound
on the counting function \( N_{p}(V) \), estimates on eigenvalue sums can be
deduced from the identity
\begin{equation}
\label{snpnp}
S_{p}(V)=\int _{0}^{\infty }N_{p}(V-pu)du.
\end{equation}
We shall obtain the following estimates.

\begin{thm}
Assume that \( V\geq 0 \), \( V\in L^{\frac{d+1}{2}}(\mathbb R^{d})\cap L^{d+1}(\mathbb R^{d}) \)
and \( 0<M\leq p \). Then there exist finite constants \( c=c(d) \) independent
on \( V \), \( M \) and \( p \), such that
\begin{eqnarray}
S_{p}(V) & \leq  & c\left( p\left( 1+\ln pM^{-1}\right) \left\Vert V\right\Vert _{L^{2}}^{2}+p^{-1}\left\Vert V\right\Vert _{L^{4}}^{4}\right) ,\quad d=3,\label{bd24} \\
S_{p}(V) & \leq  & c\left( p^{\frac{d-1}{2}}\left\Vert V\right\Vert _{L^{\frac{d+1}{2}}}^{\frac{d+1}{2}}+p^{-1}\left\Vert V\right\Vert _{L^{d+1}}^{d+1}\right) ,\quad d\geq 4.\label{bd245} 
\end{eqnarray}

\end{thm}
\begin{rem}
The respective asymptotics in section 4 show that the r.h.s. of (\ref{bd245})
captures the correct asymptotical order of the phase space average \( \Sigma _{p}(V) \)
as \( p\to \infty  \), while (\ref{bd24}) carries an additional logarithmic
factor similar to (\ref{bd13}).
\end{rem}
\begin{thm}
Assume that \( d\geq 3 \), \( V\geq 0 \) and \( V\in L^{\theta }_{w}(\mathbb {R}^{d})\cap L^{d}(\mathbb {R}^{d}) \)
for \( \frac{d+1}{2}<\theta <\frac{d}{2}+1 \). Then there exist finite constants
\( c_{1}(\theta ) \) and \( c_{2}(\theta ) \) independent on \( p \), \( M \)
and \( V \), such that
\begin{equation}
\label{bd24theta}
S_{p}(V)\leq c_{1}(\theta )p^{\theta -1}M^{d+1-2\theta }\left\Vert V\right\Vert ^{\theta }_{\theta ,w}+c_{2}(\theta )\left\Vert V\right\Vert _{L^{d+1}}^{d+1}
\end{equation}
for all \( 0<M\leq p \).
\end{thm}
\begin{rem}
The asymptotics in section 4 show that the r.h.s. of the second estimate has
the same asymptotical order in \( p \) as \( \Sigma _{p}(V) \) for \( p\to \infty  \),
if the potential \( V\geq 0 \) satisfies \( \delta _{\theta }(V)=\Delta _{\theta }(V)=v>0 \).
\end{rem}

\subsection{Potentials \protect\( V\in L^{\frac{d+1}{2}}(\mathbb {R}^{d})\cap L^{d+1}(\mathbb {R}^{d})\protect \).}

First put \( d=3 \). Standard variational arguments and the Aizenman-Lieb integration
\cite{AL} of the bound (\ref{bd13}) give
\begin{eqnarray*}
S_{p}(V) & \leq  & \co p\left( 1+\ln pM^{-1}\right) \int _{0}^{\infty }du\int _{\mathbb R^{d}}(V-pu)_{+}dx\\
 &  & \qquad +\co \int _{0}^{\infty }du\int _{\mathbb R^{d}}(V-pu)_{+}^{3}dx,
\end{eqnarray*}
which implies (\ref{bd24}). In higher dimensions a similar integration of (\ref{bd13d4})
implies (\ref{bd245}).

\subsection{Potentials \protect\( V_{+}\in L^{\theta }_{w}(\mathbb {R}^{d})\cap L^{d+1}(\mathbb {R}^{d})\protect \)
with \protect\( \frac{d+1}{2}<\theta <\frac{d}{2}+1\protect \).}

The inequality (\ref{the11}) contains a term with a weak \( L_{w}^{\theta } \)-norm.
In contrast to the usual \( L^{p} \)-norms, these weak norms in the bound for
the counting function cannot be carried over a respective weak norm in the Lieb-Thirring
inequality via the Aizenman-Lieb trick. In fact, for the proof of our results
below it shows to be necessary to refine (\ref{the11}) for potentials \( V=(W-pu)_{+} \).

Using the same notation as in the previous section in analogy to (\ref{v3co})
we first find that
\begin{equation}
\label{v3co0}
\nu _{q_{p},3}(s)\leq \co s^{-2d}\int _{\mathbb R^{d}}(W(x)-pu)_{+}^{d}dx.
\end{equation}
On the other hand, in analogy to (\ref{OO}) passing to the integration of rearrangements
we find
\begin{eqnarray*}
\nu _{q_{p},1}+\nu _{q_{p},2} & \leq  & \co \int _{\Omega _{1}\cup \Omega _{2}}\min \left\{ \frac{p^{\frac{d}{2}+1}}{Ms^{d}}(W-pu)_{+}^{\frac{d}{2}},\frac{p^{\frac{d+1}{2}}}{s^{d-1}}(W-pu)^{\frac{d-1}{2}}_{+}\right\} dx\\
 & \leq  & \co \int _{0}^{\infty }\min \left\{ \frac{p^{\frac{d}{2}+1}}{Ms^{d}}(W^{*}-pu)_{+}^{\frac{d}{2}},\frac{p^{\frac{d+1}{2}}}{s^{d-1}}(W^{*}-pu)^{\frac{d-1}{2}}_{+}\right\} dt.
\end{eqnarray*}
Put \( W\in L_{w}^{\theta } \) and \( \left\Vert W\right\Vert _{w,\theta }\leq v \),
that is \( W^{*}(t)\leq vt^{-\frac{1}{\theta }} \) for \( t>0 \). Then we
see that
\begin{eqnarray*}
\nu _{q_{p},1}(s)+\nu _{q_{p},2}(s) & \leq  & \co \frac{p^{\frac{d}{2}+1}}{Ms^{d}}\int _{t_{c}}^{\infty }(vt^{-\frac{1}{\theta }}-pu)_{+}^{\frac{d}{2}}dt\\
 &  & \qquad +\co \frac{p^{\frac{d+1}{2}}}{s^{d-1}}\int _{0}^{t_{c}}(vt^{-\frac{1}{\theta }}-pu)^{\frac{d-1}{2}}_{+}dt,
\end{eqnarray*}
where \( t_{c}=v^{\theta }(pu+p^{-1}s^{2}M^{2})^{-\theta } \). The later integral
transforms into
\begin{eqnarray}
\nu _{q_{p},1}(s)+\nu _{q_{p},2}(s) & \leq  & \notag \co \frac{v^{\theta }p^{\frac{d}{2}+1}}{Ms^{d}}\int ^{p^{-1}s^{2}M^{2}}_{0}(t+pu)^{-\theta -1}t^{\frac{d}{2}}dt\\
 &  & \qquad +\co \frac{v^{\theta }p^{\frac{d+1}{2}}}{s^{d-1}}\int _{p^{-1}s^{2}M^{2}}^{\infty }(t+pu)^{-\theta -1}t^{\frac{d-1}{2}}dt.\label{lit} 
\end{eqnarray}
Notice that for \( \frac{d+1}{2}<\theta <\frac{d}{2}+1 \) we have
\begin{eqnarray}
\int _{0}^{a}(t+\tilde{u})^{-\theta -1}t^{\frac{d}{2}}dt & \leq  & \co \min \left\{ a^{\frac{d}{2}+1}\tilde{u}^{-\theta -1},\tilde{u}^{\frac{d}{2}-\theta }\right\} ,\label{lit1} \\
\int _{a}^{\infty }(t+\tilde{u})^{-\theta -1}t^{\frac{d-1}{2}}dt & \leq  & \co \min \left\{ \tilde{u}^{\frac{d-1}{2}-\theta },a^{\frac{d-1}{2}-\theta }\right\} ,\label{lit12} 
\end{eqnarray}
where the minimum is taken for the first elements of the respective sets if
\( 0<a\leq \tilde{u} \), and for the second elements if \( 0<\tilde{u}\leq a \).
From (\ref{lit}) and (\ref{lit1}), (\ref{lit12}) we conclude that
\begin{eqnarray*}
\nu _{q_{p},1}(s)+\nu _{q_{p},2}(s) & \leq  & \co v^{\theta }\left( s^{2}M^{d+1}p^{-\theta -1}u^{-\theta -1}+s^{1-d}p^{d-\theta }u^{\frac{d-1}{2}-\theta }\right) \\
 & \leq  & \co v^{\theta }s^{1-d}p^{d-\theta }u^{\frac{d-1}{2}-\theta }
\end{eqnarray*}
if \( s^{2}M^{2}p^{-2}\leq u \), and
\[
\nu _{q_{p},1}(s)+\nu _{q_{p},2}(s)\leq \co v^{\theta }\left( p^{d+1-\theta }M^{-1}s^{-d}u^{\frac{d}{2}-\theta }+p^{1+\theta }M^{d-1-2\theta }s^{-2\theta }\right) \]
for \( s^{2}M^{2}p^{-2}\geq u \). These two bounds in conjunction with (\ref{v3co0})
give
\begin{eqnarray*}
\nu _{q_{p}}(s) & \leq  & \co s^{-2d}\left\Vert (W-pu)_{+}\right\Vert ^{d}_{L^{d}}+\co v^{\theta }\min \left\{ s^{1-d}p^{d-\theta }u^{\frac{d-1}{2}-\theta },\right. \\
 &  & \qquad \left. p^{d+1-\theta }M^{-1}s^{-d}u^{\frac{d}{2}-\theta }+p^{1+\theta }M^{d-1-2\theta }s^{-2\theta }\right\} ,\quad s>0.
\end{eqnarray*}
The inverse \( q^{*}_{p} \) of \( \nu _{q_{p}} \) satisfies then the bound
\begin{eqnarray*}
q^{*}_{p}(t) & \leq  & \co t^{-\frac{1}{2d}}\left\Vert (W-pu)_{+}\right\Vert ^{d}_{L^{d}}+\co \min \left\{ t^{\frac{1}{1-d}}v^{\frac{\theta }{d-1}}p^{\frac{d-\theta }{d-1}}u^{\frac{1}{2}-\frac{\theta }{d-1}},\right. \\
 &  & \qquad \left. t^{-\frac{1}{d}}v^{\frac{\theta }{d}}p^{1+\frac{1-\theta }{d}}M^{-\frac{1}{d}}u^{\frac{1}{2}-\frac{\theta }{d}}+t^{-\frac{1}{2\theta }}v^{\frac{1}{2}}p^{\frac{1+\theta }{2\theta }}M^{\frac{d-1}{2\theta }-1}\right\} 
\end{eqnarray*}
for all \( t>0 \). Hence, if \( d\geq 4 \) we get
\begin{eqnarray*}
\left\langle q_{p}\right\rangle (t) & \leq  & \co t^{-\frac{1}{2d}}\left\Vert (W-pu)_{+}\right\Vert ^{d}_{L^{d}}+\co \min \left\{ \frac{v^{\frac{\theta }{d-1}}p^{\frac{d-\theta }{d-1}}}{t^{\frac{1}{d-1}}u^{\frac{\theta }{d-1}-\frac{1}{2}}},\right. \\
 &  & \qquad \left. \frac{v^{\frac{\theta }{d}}p^{1+\frac{1-\theta }{d}}u^{\frac{1}{2}-\frac{\theta }{d}}}{t^{\frac{1}{d}}M^{\frac{1}{d}}}+\frac{v^{\frac{1}{2}}p^{\frac{1+\theta }{2\theta }}M^{\frac{d-1}{2\theta }-1}}{t^{\frac{1}{2\theta }}}\right\} 
\end{eqnarray*}
for all \( t>0 \), while for the dimension \( d=3 \) we obtain
\begin{eqnarray*}
\left\langle q_{p}\right\rangle (t) & \leq  & \co t^{-\frac{1}{6}}\left\Vert (W-pu)_{+}\right\Vert ^{\frac{1}{2}}_{L^{3}}+\co \min \left\{ \frac{v^{\frac{1}{2}}p^{\frac{1}{2\theta }+\frac{1}{2}}}{t^{\frac{1}{2\theta }}M^{1-\frac{1}{\theta }}}+\right. \\
 &  & \left. +\frac{v^{\frac{\theta }{3}}p^{\frac{4-\theta }{3}}u^{\frac{1}{2}-\frac{\theta }{3}}}{t^{\frac{1}{3}}M^{\frac{1}{3}}},\frac{v^{\frac{\theta }{2}}p^{\frac{3-\theta }{2}}u^{\frac{1-\theta }{2}}}{t^{\frac{1}{2}}}\left( 1+\ln _{+}\left( \frac{tu^{\theta }p^{\theta -1}}{v^{\theta }M^{2}}\right) \right) \right\} 
\end{eqnarray*}
as \( t>0 \). In view of (\ref{BS}) we conclude, that it holds either in higher
dimesions
\begin{eqnarray}
N_{p}(W-pu) & \leq  & \notag \co \left\Vert (W-pu)_{+}\right\Vert ^{d}_{L^{d}}+\co v^{\theta }\min \left\{ p^{d-\theta }u^{\frac{d-1}{2}-\theta },\right. \\
 &  & \qquad \left. p^{d+1-\theta }M^{-1}u^{\frac{d}{2}-\theta }+p^{1+\theta }M^{d-1-2\theta }\right\} ,\quad d\geq 4,\label{np0} 
\end{eqnarray}
or
\begin{eqnarray}
N_{p}(V-pu) & \leq  & \co \left\Vert (W-pu)_{+}\right\Vert ^{d}_{L^{d}}+\co v^{\theta }\min \left\{ M^{-1}p^{4-\theta }u^{\frac{3}{2}-\theta }+\right. \notag \\
 &  & \left. +M^{2-2\theta }p^{1+\theta },u^{-\theta }M^{2}p^{1-\theta }f(C\sqrt{u}pM^{-1})\right\} \quad \mbox {if}\quad d=3,\label{np} 
\end{eqnarray}
where \( C \) is some fixed finite positive constant and \( y=f(x) \) is the
inverse function to \( x=\sqrt{y}/(1+\ln _{+}Cy) \) on \( \mathbb {R}_{+} \). 

Inserting (\ref{np0}) into (\ref{snpnp}) we obtain immediately (\ref{bd24theta})
for \( d\geq 4 \). To settle the case \( d=3 \) we first note that \( f(x)\leq cx^{2}(1+\ln _{+}\sqrt{C}x)^{2} \)
if \( c \) is chosen such that \( \sqrt{c}\geq (1+\ln _{+}t)\left( 1+\ln _{+}\frac{t}{1+\ln _{+}t}\right) ^{-1} \)
for all \( t>0 \). Hence, the bound (\ref{np}) can be developed as follows
\begin{equation}
\label{np1}
N_{p}(V-pu)\leq \co \left\{ \begin{array}{ccc}
v^{\theta }M^{-1}p^{4-\theta }u^{\frac{3}{2}-\theta }+v^{\theta }M^{2-2\theta }p^{1+\theta } & \: \mbox {for}\:  & u\leq \frac{M^{2}}{Cp^{2}}\\
v^{\theta }p^{3-\theta }u^{1-\theta }(1+\ln _{+}(\sqrt{Cu}pM^{-1}))^{2} & \: \mbox {for}\:  & u>\frac{M^{2}}{Cp^{2}}
\end{array}\right. .
\end{equation}
For \( \theta >2 \), the following identity holds true
\begin{eqnarray}
 &  & \int _{a^{-2}}^{\infty }u^{1-\theta }(1+\ln (a\sqrt{u}))^{2}du\notag \\
 & = & \left( \frac{1}{\theta -2}+\frac{1}{(\theta -2)^{2}}+\frac{1}{2(\theta -2)^{3}}\right) a^{2\theta -4}\label{np2} 
\end{eqnarray}
for any \( a>0 \). If we integrate (\ref{np1}) in \( u \) for \( 2<\theta <\frac{5}{2} \)
and take (\ref{np2}) into account, we arrive at (\ref{bd24theta}).

\section{Asymptotics of the Eigenvalue Moments and the Counting Function}

\subsection{Statement of the main results.}

We turn now to the calculation of the asymptotical behaviour of \( \Sigma _{p}(V) \)
and \( N_{p}(V) \) for certain cases. In particular, we shall obtain the following
two formulae:

\begin{thm}
\label{sec: maintm1}Assume that \( V\in L^{\theta }(\mathbb {R}^{3})\cap L^{4}(\mathbb {R}^{3}) \)
for some \( \theta <2 \) and that \( V \) has uniformly bounded, continuous
second derivatives if \( d=3 \), or that \( V\in L^{\frac{d+1}{2}}(\mathbb {R}^{d})\cap L^{d+1}(\mathbb {R}^{d}) \)
if \( d\geq 4 \). Then the asymptotical formula
\begin{equation}
\label{mainas1}
S_{p}(V)=(1+o(1))\Sigma _{p}(V)=\frac{(1+o(1))p^{\frac{d-1}{2}}\omega _{d}}{(d+1)2^{\frac{3d-1}{2}}\pi ^{d}}\int _{\mathbb {R}^{d}}V_{+}^{\frac{d+1}{2}}(y)dy
\end{equation}
holds true as \( p\to \infty  \).
\end{thm}
\begin{rem}
For \( d=3 \) the assumptions on the potential \( V \) in Theorem \ref{sec: maintm1}
are more restrictive than the natural one \( V\in L^{2}\cap L^{4} \). The additional
logarithmic factor in (\ref{bd24}) prevents one to use this bound to close
formula (\ref{mainas1}) to the natural class of potentials. It remains an open
problem, whether (\ref{mainas1}) holds actually for all \( V\in L^{2}\cap L^{4} \)
if \( d=3 \).
\end{rem}
\begin{thm}
\label{sec: maintm2}Assume that \( U,V\geq 0 \), \( U,V\in L^{\theta }\cap L^{d+1} \)
for some \( \theta <\frac{d+1}{2} \) and that \( U \) and \( V \) possess
uniformly bounded second derivatives. Put \( U(y;p)=U(p^{-1}y) \). Then
\begin{equation}
\label{mainas2}
\lim _{p\to \infty }p^{-\frac{d+1}{2}}tr\: U(y;p)\chi _{0}(Q_{p}(i\nabla ,y))=\frac{\omega _{d}}{2^{\frac{3d+1}{2}}\pi ^{d}}\int U(x)V^{\frac{d-1}{2}}(x)dx.
\end{equation}

\end{thm}
We mention the following obvious consequence of Theorem \ref{sec: maintm2}:

\begin{cor}
If \( V\geq 0 \) has uniformly bounded second derivatives and \( V\in L^{\frac{d-1}{2}}\cap L^{d+1} \)
then
\[
\liminf _{p\to \infty }p^{-\frac{d+1}{2}}N_{p}(V)\geq \frac{\omega _{d}}{2^{\frac{3d+1}{2}}\pi ^{d}}\int V^{\frac{d-1}{2}}dx.\]

\end{cor}
The remaining part of this paper is devoted to the proof of Theorem \ref{sec: maintm1}
and Theorem \ref{sec: maintm2}. Our approach is based on the methods of coherent
states. Therefore we first give a short survey of the necessary general material
from this subject.

\subsection{Coherent States and Berezin-Lieb Inequalities: Preliminaries.}

Fix some spherically symmetric, smooth, non-negative function \( f \) with
compact support in \( \mathbb {R}^{d} \), such that \( \left\Vert f\right\Vert _{L^{2}(\mathbb {R}^{d})}=1 \).
Put \( f_{\epsilon }(x)=\epsilon ^{d/2}f(\epsilon x) \) where \( \epsilon >0 \).
For given \( \gamma =\{y,\xi \} \) with \( y,\xi \in \mathbb {R}^{d} \) we
define the coherent states
\begin{equation}
\label{CoHe}
\Pi _{\gamma }^{\epsilon }(x)=e^{-i\xi x}f_{\epsilon }(x-y).
\end{equation}
For any fixed \( \gamma  \) and \( \epsilon  \) it holds \( \left\Vert \Pi _{\gamma }^{\epsilon }\right\Vert _{L^{2}(\mathbb {R}^{d})}=1 \). 

Let \( J \) be a non-negative, locally integrable function on \( \mathbb {R}^{d} \)
with not more than polynomial growth at infinity. We define the operator \( J(i\nabla )=\Phi ^{*}J\Phi  \)
in the usual way with \( \Phi  \) being the unitary Fourier transformation.
Put \( \hat{f}=\Phi f \). In view of our choice of coherent states it is associated
with the symbol function
\begin{equation}
\label{jint}
j_{\epsilon }(\gamma )=j_{\epsilon }(\xi )=(J(i\nabla _{x})\Pi _{\gamma }^{\epsilon }(x),\Pi ^{\epsilon }_{\gamma }(x))_{L^{2}(\mathbb R^{d},dx)}=(J\star |\hat{f}_{\epsilon }|^{2})(\xi ).
\end{equation}
The operator of multiplication by a locally integrable real-valued function
\( W \) on \( \mathbb {R}^{3} \) corresponds to the symbol
\[
w_{\epsilon }(\gamma )=w_{\epsilon }(y)=(W(x)\Pi _{\gamma }^{\epsilon }(x),\Pi ^{\epsilon }_{\gamma }(x))_{L^{2}(\mathbb R^{d},dx)}=(W\star f_{\epsilon }^{2})(y),\]
Here \( (\cdot ,\cdot )_{L^{2}(\mathbb R^{d},dx)} \) is the scalar product
in \( L^{2}(\mathbb R^{d}) \) with respect to the variable \( x \) and \( u\star v \)
denotes the convolution
\[
(u\star v)(x)=\int u(x-x^{\prime })v(x^{\prime })dx^{\prime }.\]
 If now \( W=W_{1}+W_{2} \), where \( W_{1} \) is uniformly bounded and \( W_{2} \)
is form compact with respect to \( J(i\nabla ) \), the operator sum \( J(i\nabla )+W(x) \)
can be defined in the form sense. Let \( \psi  \) be some non-negative convex
function on \( \mathbb {R} \), such that \( \psi (J(i\nabla )+W(x)) \) is
trace class. Then the Lieb-Berezin inequality states that (\cite{Be}, see also
\cite{LS})
\begin{equation}
\label{LB1}
\int _{\mathbb {R}^{2d}}\psi (j_{\epsilon }(\xi )+w_{\epsilon }(y))d\gamma \leq tr\, \psi (J(i\nabla )+W(x)).
\end{equation}
Moreover, if the average of \( \psi (J(\xi )+W(y)) \) in \( \mathbb {R}^{2d} \)
with respect to \( d\gamma  \) is finite, then \( \psi (j_{\epsilon }(i\nabla )+w_{\epsilon }(x)) \)
is trace class and
\begin{equation}
\label{LB2}
tr\, \psi (j_{\epsilon }(i\nabla )+w_{\epsilon }(x))\leq \int _{\mathbb {R}^{2d}}\psi (J(\xi )+W(y))d\gamma 
\end{equation}
 Let us finally assume that in addition to this \( J \) or \( W \) are twice
continuously differentiable with the following uniform bounds on the matrix
norms of the respective Hessians
\[
\vartheta (J)=\max _{\xi \in \mathbb {R}^{d}}\left\Vert \left\{ \frac{\partial ^{2}J}{\partial \xi _{l}\partial \xi _{k}}\right\} _{k,l=1}^{d}\right\Vert \quad \mbox {and}\quad \vartheta (W)=\max _{\xi \in \mathbb {R}^{d}}\left\Vert \left\{ \frac{\partial ^{2}W}{\partial \xi _{l}\partial \xi _{k}}\right\} _{k,l=1}^{d}\right\Vert .\]
Put
\[
\psi (x)=x_{-}=\left\{ \begin{array}{lcl}
-x & \quad \mbox {for}\quad  & x<0\\
0 & \quad \mbox {for}\quad  & x\geq 0
\end{array}\right. .\]
We also recall that \( \chi _{-t} \) is the characteristic function of the
interval \( (-\infty ,-t) \). Under the above conditions we have

\begin{lem}
\label{sec: Lme1}The two-sided bound
\begin{eqnarray}
\int (J(\xi )+W(y)+\kappa )_{-}d\gamma  & \leq  & tr\, (J(i\nabla )+W(x))_{-},\label{bdulm1} \\
tr\, (J(i\nabla )+W(x))_{-} & \leq  & \int (J(\xi )+W(y))_{-}d\gamma +\Theta _{\kappa }\label{bdolm1} 
\end{eqnarray}
holds true, where
\[
\kappa =2\sqrt{\vartheta (J)\vartheta (W)}\left\Vert xf(x)\right\Vert \left\Vert \nabla f\right\Vert \]
and
\[
\Theta _{\kappa }=\int _{0}^{\kappa }tr\: \chi _{-t}(J(i\nabla )+W(x))dt.\]

\end{lem}
\begin{proof}
Indeed, by Taylors formula we have
\[
J(\xi -\xi ^{\prime })=J(\xi )-\xi ^{\prime }\cdot \nabla J(\xi )+\sum _{k,l}\frac{\partial ^{2}J(\tilde{\xi }(\xi ,\xi ^{\prime }))}{\partial \xi _{k}\partial \xi _{l}}\xi ^{\prime }_{k}\xi ^{\prime }_{l},\]
where \( \tilde{\xi } \) is some point on the line segment connecting \( \xi  \)
and \( \xi ^{\prime } \). Inserting this into the integral expression for (\ref{jint}),
because of \( \left\Vert \hat{f}_{\epsilon }\right\Vert _{L^{2}(\mathbb {R}^{d})}=1 \)
one finds that
\[
j_{\epsilon }(\xi )-J(\xi )=-\nabla J(\xi )\cdot \int \xi ^{\prime }|\hat{f}_{\epsilon }(\xi ^{\prime })|^{2}d\xi ^{\prime }+\sum _{k,l}\int \frac{\partial ^{2}J(\tilde{\xi })}{\partial \xi _{k}\partial \xi _{l}}\xi ^{\prime }_{k}\xi ^{\prime }_{l}|\hat{f}_{\epsilon }(\xi ^{\prime })|^{2}d\xi ^{\prime }.\]
Since \( \hat{f}_{\epsilon } \) is spherically symmetric, the first integral
on the r.h.s. vanishes and
\begin{eqnarray}
\notag |j_{\epsilon }(\xi )-J(\xi )| & \leq  & \vartheta (J)\int |\xi ^{\prime }|^{2}|\hat{f}_{\epsilon }(\xi ^{\prime })|^{2}_{L^{2}(\mathbb {R}^{d})}d\xi ^{\prime }\\
 & \leq  & \vartheta (J)\epsilon ^{2}\left\Vert \nabla f\right\Vert ^{2}_{L^{2}(\mathbb {R}^{d})}.\label{erj} 
\end{eqnarray}
In a similarly way we get
\begin{equation}
\label{erw}
|w_{\epsilon }(y)-W(y)|\leq \vartheta (W)\epsilon ^{-2}\left\Vert xf(x)\right\Vert ^{2}_{L^{2}(\mathbb {R}^{d})}.
\end{equation}
Now (\ref{LB1}), (\ref{erj}) and (\ref{erw}) for the optimal choice of \( \epsilon  \)
give the first inequality of Lemma \ref{sec: Lme1}. On the other hand (\ref{erj})
and (\ref{erw}) imply
\[
J(i\nabla )+W(x)+\kappa \geq j_{\epsilon _{0}}(i\nabla )+w_{\epsilon _{0}}(x)\]
and
\[
tr\, (J(i\nabla )+W(x))_{-}\leq tr\, (j_{\epsilon _{0}}(i\nabla )+w_{\epsilon _{0}}(x))_{-}+tr\: g_{\kappa }(J(i\nabla )+W(x))\]
with \( g_{\kappa }(x)=\min \left\{ \kappa ,-x\right\}  \) for \( x<0 \) and
\( g_{\kappa }(x)=0 \) for \( x\geq 0 \). Since
\[
g_{\kappa }(x)=\int _{0}^{\kappa }\chi _{-t}(x)dt,\]
the bound (\ref{LB2}) implies the second statement of the Lemma.
\end{proof}

\section{Moments of negative Eigenvalues. An Estimate from Below.}

\subsection{Summary.}

We turn here to the study of the asymptotics of eigenvalue moments
\[
S(p)=tr\, (Q_{p}(i\nabla ,y))_{-},\quad Q_{p}(i\nabla ,y)=H_{p}(\xi )-V_{p}(y).\]
Because of the divergence of the second derivatives of \( H_{p}(\xi ) \) near
the points \( e_{\pm }=(\pm \mu _{\pm },0,\dots ,0)\in \mathbb R^{d} \) as
\( p\to \infty  \), a straightforward application of the bound (\ref{bdulm1})
in Lemma \ref{sec: Lme1} will not lead to the desired results. Therefore we
have to implement a suitable smoothing procedure of the symbol first. In this
section we consider the bound from below.

\subsection{Basic properties of the symbol \protect\( H_{p}(\xi )\protect \)}

Consider the functions
\[
T_{\pm }(\xi )=\sqrt{(\eta \mp \mu _{\pm })^{2}+|\zeta |^{2}+\mu ^{2}_{\pm }M^{2}p^{-2},}\]
with \( \xi \in {\Bbb R}^{d} \), \( \xi =(\eta ,\zeta ) \) for \( \xi _{1}=\eta \in {\Bbb R} \)
and \( (\xi _{2},\dots ,\xi _{d})=\zeta \in {\Bbb R}^{d-1} \), \( M=m_{+}+m_{-} \),
\( \mu _{\pm }=m_{\pm }M^{-1} \). Here \( m_{\pm } \) and \( p \) are positive
parameters. We have
\[
H_{p}(\xi )=T_{+}(\xi )+T_{-}(\xi )-\sqrt{1+M^{2}p^{-2}}.\]
This is a convex non-negative function, which is rotational symmetric with respect
to the \( \eta  \)-axes. It achieves a unique, non-degenerate minimum at the
point \( \xi =0 \) where \( H_{p}(0)=0 \). 

The gradient and the Hessian of \( T_{\pm } \) calculate as follows
\begin{eqnarray*}
\nabla T_{\pm }(\xi ) & = & T_{\pm }^{-1}(\xi )\left( \eta \mp \mu _{\pm },\zeta ^{t}\right) ^{t},\\
(\nabla \nabla ^{t})T_{\pm } & = & T_{\pm }^{-1}\left( {\Bbb I}-(\nabla T_{\pm })(\nabla T_{\pm })^{t}\right) .
\end{eqnarray*}
Hence,
\begin{equation}
\label{dh}
\left| \frac{\partial H_{p}(\xi )}{\partial \xi _{k}}\right| \leq 2\quad \mbox {and}\quad \left| \frac{\partial ^{2}H_{p}(\xi )}{\partial \xi _{k}\partial \xi _{l}}\right| \leq T_{+}^{-1}(\xi )+T_{-}^{-1}(\xi )
\end{equation}
for all \( \xi \in \mathbb {R}^{d} \), \( p,M>0 \) and \( l,k=1,\dots ,d \).

\subsection{Smoothing of the symbol.}

Let \( g \) be a smooth, spherically symmetric non-negative function on \( {\Bbb R}^{d} \)
supported within the unit ball, such that \( \int g(x)dx=1 \). If \( \sigma >0 \)
we put \( g_{\sigma }(x)=\sigma ^{-d}g(\sigma ^{-1}x) \), for \( \sigma =0 \)
we set \( g_{0}(x)=\delta (\cdot -x) \) and define
\begin{eqnarray}
H_{p,\sigma }(\xi ) & = & \int _{\mathbb R^{d}}H_{p}(\xi -y)g_{\sigma (\xi )}(y)dy\label{hg} \\
\notag  & = & \int _{\mathbb R^{d}}H_{p}(\xi -\sigma (\xi )t)g(t)dt.
\end{eqnarray}
It holds

\begin{lem}
\label{sec: pointwise}The functions \( H_{p}(\xi ) \) and \( H_{p,\sigma }(\xi ) \)
satisfy the pointwise estimate
\begin{equation}
\label{hphg}
H_{p}(\xi )\leq H_{p,\sigma }(\xi ),\quad \xi \in \mathbb R^{d}.
\end{equation}

\end{lem}
\begin{proof}
Note that \( H_{p} \) is convex and the spherically symmetric weight \( g_{\sigma } \)
has the total mass \( 1 \). If we represent in (\ref{hg}) the term \( H_{p}(\xi -y) \)
in a Taylor series at the point \( \xi  \) of order one with a positive quadratic
form as remainder term, the inequality (\ref{hphg}) follows immediately. 
\end{proof}
Put \( \tau _{\pm }=\tau _{\pm }(\xi )=|\xi -e_{\pm }| \). Below we chose
\begin{equation}
\label{sr}
\sigma (\xi )=\sigma _{r}(\xi )=\left\{ \begin{array}{lcl}
0 & \mbox {if} & \xi \not \in B_{r}^{+}\cup B_{r}^{-}\\
re^{\varsigma _{-}(\xi ,r)} & \mbox {if} & \xi \in B^{-}_{r}\\
re^{\varsigma _{+}(\xi ,r)} & \mbox {if} & \xi \in B_{r}^{+}
\end{array}\right. ,
\end{equation}
where \( 0<r<\min \{\mu _{+},\mu _{-}\}/2 \), \( B_{r}^{\pm }=\{\xi :\tau _{\pm }(\xi )<r\} \)
and
\[
\varsigma _{\pm }(\xi ,r)=\frac{-1}{1-r^{-2}\tau _{\pm }^{2}(\xi )}.\]

\begin{lem}
\label{sec: derivs}One can find an appropriate finite constant \( c \), which
is independent on \( p,M,r>0 \), \( \xi \in \mathbb R^{d} \) and \( k,l=1,\dots \, d \),
such that
\begin{eqnarray}
\left| \frac{\partial H_{p,\sigma }}{\partial \xi _{k}}\right|  & \leq  & c,\label{dhp1} \\
\left| \frac{\partial ^{2}H_{p,\sigma }}{\partial \xi _{k}\partial \xi _{l}}\right|  & \leq  & c(1+r^{-1}).\label{dhp2} 
\end{eqnarray}

\end{lem}
\begin{proof}
Obviously it holds
\begin{equation}
\label{difH1}
\frac{\partial H_{p,\sigma }(\xi )}{\partial \xi _{k}}=\int \sum ^{d}_{j=1}\frac{\partial \nu _{j}}{\partial \xi _{k}}\frac{\partial H_{p}(\nu )}{\partial \nu _{j}}g(t)dt,\quad \nu _{j}=\xi _{j}-\sigma (\xi )t_{j},
\end{equation}
and
\begin{equation}
\label{diff}
\frac{\partial ^{2}H_{p,\sigma }(\xi )}{\partial \xi _{k}\partial \xi _{l}}=\int \left\{ \sum _{j=1}^{d}\frac{\partial ^{2}\nu _{j}}{\partial \xi _{k}\partial \xi _{l}}\frac{\partial H_{p}(\nu )}{\partial \nu _{j}}+\sum _{j,i=1}^{d}\frac{\partial \nu _{j}}{\partial \xi _{k}}\frac{\partial \nu _{i}}{\partial \xi _{l}}\frac{\partial ^{2}H_{p}(\nu )}{\partial \nu _{j}\partial \nu _{i}}\right\} g(t)dt.
\end{equation}
Since
\begin{equation}
\label{difg}
\left| \frac{\partial \sigma _{r}}{\partial \xi _{k}}\right| \leq \co \quad \mbox {and}\quad \left| \frac{\partial ^{2}\sigma _{r}}{\partial \xi _{k}\partial \xi _{l}}\right| \leq \co r^{-1},
\end{equation}
from (\ref{difH1}) and the first estimates in (\ref{dh}), (\ref{difg}) we
conclude (\ref{dhp1}).

To estimate the second derivatives we note , that by (\ref{dh}) and (\ref{difg})
the first part of the integral on the r.h.s. of (\ref{diff}) can be estimated
by \( \co (1+r^{-1}) \), while the second term in (\ref{diff}) does not exceed
\[
\co \int (T^{-1}_{+}(\nu )+T^{-1}_{-}(\nu ))g(t)dt.\]
Note that \( T_{\pm }(\nu )\geq |\tau _{\pm }-\sigma _{r}t| \) and because
\( g \) is bounded and of compact support we have
\begin{eqnarray*}
\int _{\mathbb R^{d}}\frac{g(t)dt}{T_{\pm }(\nu )} & \leq  & \co \frac{\tau _{\pm }^{d-1}}{\sigma _{r}^{d}}\int _{\mathbb S^{d-2}}d\phi \int _{0}^{\pi }d\theta \sin \theta \int _{0}^{\frac{\sigma _{r}}{\tau _{\pm }}}\frac{t^{d-1}dt}{\sqrt{1+t^{2}-2t\cos \theta }}\\
 & \leq  & \co \frac{\tau _{\pm }^{d-1}}{\sigma _{r}^{d}}\int _{0}^{\frac{\sigma _{r}}{\tau _{\pm }}}t^{d-2}(t+1-|t-1|)dt\\
 & \leq  & \co \min \{\tau _{\pm }^{-1},\sigma _{r}^{-1}\}.
\end{eqnarray*}
For \( 0\leq \tau _{\pm }\leq r/2 \) the function \( \sigma _{r} \) can be
estimated by \( \sigma _{r}\geq e^{-4/3}r \). Hence,
\[
\int \frac{g(t)dt}{T_{\pm }(\nu )}\leq c(1+r^{-1})\]
and we conclude (\ref{dhp2}).
\end{proof}

\subsection{The estimate from below.}

We are now in the position to obtain the main result of this section. Put
\begin{eqnarray*}
Q_{p,\sigma _{r}}(\xi ,y) & = & H_{p,\sigma _{r}}(\xi )-V_{p}(y),\\
Q_{p,\sigma _{r}}(i\nabla ,y) & = & H_{p,\sigma _{r}}(i\nabla )-V_{p}(y).
\end{eqnarray*}
By Lemma \ref{sec: pointwise} we find that
\begin{equation}
\label{st1}
tr\, (Q_{p}(i\nabla ,y))_{-}\geq tr\, (Q_{p,\sigma _{r}}(i\nabla ,y))_{-}.
\end{equation}
Next we apply the first part of Lemma \ref{sec: Lme1} with \( J=H_{p,\sigma _{r}} \)
and \( W=V_{p} \) to this bound. By (\ref{dhp2}) we have \( \vartheta (H_{p,\sigma _{r}})\leq cr^{-1} \)
for \( 0<r<\min \{\mu _{+},\mu _{-}\} \), while \( \vartheta (V_{p})\leq p^{-3}\vartheta (V) \).
Then (\ref{bdulm1}) implies that
\begin{equation}
\label{st2}
tr\, (Q_{p,\sigma _{r}}(i\nabla ,y))_{-}\geq \int (Q_{p,\sigma _{r}}(\xi ,y)+\kappa )_{-}d\gamma ,
\end{equation}
where \( \kappa \leq \co \sqrt{\vartheta (V)r^{-1}p^{-3}} \). From (\ref{st1})
and (\ref{st2}) we finally conclude

\begin{lem}
\label{lm: esfrbl}The inequality
\begin{equation}
\label{st3}
S(p)\geq \int (Q_{p,\sigma _{r}}(\xi ,y)+\kappa )_{-}d\gamma 
\end{equation}
holds true for some \( \kappa \leq c\sqrt{\vartheta (V)r^{-1}p^{-3}} \), where
the constant \( c \) in the estimate for \( \kappa  \) can be chosen to be
independent on \( V \), \( p \), \( M \) and \( r \), \( 0<r<\min \{\mu _{+},\mu _{-}\} \).
\end{lem}

\section{Moments of Negative Eigenvalues: An Estimate from Above}

\subsection{Summary. }

We shall now accompany Lemma \ref{lm: esfrbl} by a corresponding estimates
from above. As in the previous section we smooth the symbol before applying
(\ref{bdolm1}) from Lemma \ref{sec: Lme1}. But in the absence of a replacement
of Lemma \ref{sec: pointwise} we have to modify the symbol additionally.

\subsection{Modification of the symbol.}

We put \( \delta \in (0,1/2) \), \( \xi =(\eta ,\zeta ) \) with \( \xi _{1}=\eta \in {\Bbb R} \)
and \( (\xi _{2},\dots ,\xi _{d})=\zeta \in {\Bbb R}^{d-1} \), and set
\begin{eqnarray}
G_{p,\delta }(\xi ) & = & H_{p}((1-\delta )\eta ,\zeta ),\notag \\
G_{p,\delta ,\sigma }(\xi ) & = & \int _{\mathbb R^{d}}G_{p,\delta }(\xi -y)g_{\sigma (\xi )}(y)dy\notag \\
 & = & \int _{\mathbb R^{d}}G_{p,\delta }(\xi -\sigma (\xi )t)g(t)dt.\label{gsigma} 
\end{eqnarray}
In analogy to (\ref{sr}) let the function \( \sigma (\xi )=\sigma _{r,\delta }(\xi ) \)
be given by
\begin{equation}
\label{srd}
\sigma (\xi )=\sigma _{r,\delta }(\xi )=\left\{ \begin{array}{lcl}
0 & \mbox {if} & \xi \not \in B_{r,\delta }^{+}\cup B_{r,\delta }^{-}\\
re^{\varsigma _{-,\delta }(\xi ,r)} & \mbox {if} & \xi \in B^{-}_{r,\delta }\\
re^{\varsigma _{+,\delta }(\xi ,r)} & \mbox {if} & \xi \in B_{r,\delta }^{+}
\end{array}\right. ,
\end{equation}
where \( 0<r<\min \{\mu _{-},\mu _{+}\} \), \( B_{r,\delta }^{\pm }=\{\xi :|\xi -e_{\pm ,\delta }|<r\} \),
\( e_{\pm ,\delta }=(1-\delta )^{-1}e_{\pm } \) and
\[
\varsigma _{\pm ,\delta }(\xi ,r)=\frac{-1}{1-r^{-2}|\xi -e_{\pm ,\delta }|^{2}}.\]
Similar to the proof of Lemma \ref{sec: derivs} one can show that the derivatives
of \( G_{p,\delta ,\sigma }(\xi ) \) satisfy the bounds
\begin{eqnarray}
\left| \frac{\partial G_{p,\delta ,\sigma }}{\partial \xi _{k}}\right|  & \leq  & c,\label{dhp1g} \\
\left| \frac{\partial ^{2}G_{p,\delta ,\sigma }}{\partial \xi _{k}\partial \xi _{l}}\right|  & \leq  & c(1+r^{-1}).\label{dhp2g} 
\end{eqnarray}
The constant \( c \) in (\ref{dhp1g}), (\ref{dhp2g}) can be chosen to be
independent on \( p \), \( M \), \( r \), \( \xi  \), \( k \), \( l \),
and \( \delta \in (0,1/2) \) as well. 

\begin{lem}
There exists a finite positive constant \( C=C(\mu _{+},\mu _{-}) \), such
that the bound
\begin{equation}
\label{GH}
G_{p,\delta ,\sigma _{r}}(\xi )\leq H_{p}(\xi ),\quad \xi \in \mathbb {R}^{d},
\end{equation}
holds true for all \( r\leq \min \{\mu _{-},\mu _{+},C\delta \} \), \( 0<\delta <1/2 \)
and all \( p\geq M>0 \).
\end{lem}
\begin{proof}
Let \( r\leq \min \{\mu _{-},\mu _{+}\} \). Since
\[
G_{p,\delta ,\sigma _{r}}(\xi )=H_{p}((1-\delta )\eta ,\zeta )\quad \mbox {if}\quad \xi \not \in B_{r,\delta }^{+}\cup B_{r,\delta }^{-},\]
the bound (\ref{GH}) for that case is an obvious consequence of the local monotonicity
of \( H_{p}(\eta ,\zeta ) \) in \( \eta  \) for fixed \( p \), \( M \) and
\( \zeta  \). 

On the other hand, by (\ref{dh}) it holds \( |\partial H_{p}/\partial \eta |\leq 2 \)
and \( |\partial G_{p,\delta }/\partial \eta |\leq 2 \). Hence, if
\[
r\leq r(\delta )=(H_{p}(e_{\pm ,\delta })-G_{p}(e_{\pm ,\delta }))/4\]
we have
\begin{equation}
\label{maxest}
\min _{\xi ^{\prime }\in B_{r,\delta }^{\pm }}H_{p}(\xi ^{\prime })\geq \max _{\xi ^{\prime \prime }\in B_{r,\delta }^{\pm }}G_{p,\delta }(\xi ^{\prime \prime }).
\end{equation}
For any \( \xi \in B_{r,\delta }^{\pm } \) and \( t\in \mathbb {R}^{d} \),
\( |t|\leq 1 \) it holds
\[
|(\xi -t\sigma _{r,\delta }(\xi ))-e_{\pm ,\delta }|\leq |\xi -e_{\pm ,\delta }|+\sigma _{r,\delta }(\xi )\leq r.\]
The later inequality follows from the fact that \( x+e^{-(1-x^{2})^{-1}}\leq 1 \)
for all \( 0\leq x<1 \). Thus, the argument \( \xi ^{\prime \prime }=\xi -t\sigma _{r,\delta }(\xi ) \)
of \( G_{p,\delta } \) in (\ref{gsigma}) satisfies \( \xi ^{\prime \prime }\in B_{r,\delta }^{\pm } \)
on the support of \( g \), and we conclude (\ref{GH}) from (\ref{maxest})
and the normalisation of \( g \).

It remains to estimate \( r(\delta ) \) from below. Note that \( Mp^{-1}\leq 1 \)
and \( 0<\delta <1/2 \). Then
\begin{eqnarray*}
4r(\delta ) & = & H_{p}(e_{\pm ,\delta })-H_{p}(e_{\pm })\\
 & \geq  & \mp \frac{\delta \mu _{\pm }}{1-\delta }\min _{\mu _{\pm }\leq \eta \leq \mu _{\pm }(1-\delta )^{-1}}\frac{\partial }{\partial \eta }H_{p}(\eta ,0,0)\\
 & \geq  & \frac{\delta }{1-\delta }\frac{1}{\sqrt{\mu _{\mp }^{2}(1+(1-\delta )^{-1})^{2}+1}}\geq C(\mu _{+},\mu _{-})\delta .
\end{eqnarray*}
This completes the proof.
\end{proof}

\subsection{The estimate from above.}

We put now
\begin{eqnarray*}
Q_{p,\delta ,\sigma }(\xi ,y) & = & G_{p,\delta ,\sigma }(\xi )-V_{p}(y),\\
Q_{p,\delta ,\sigma }(i\nabla ,y) & = & G_{p,\delta ,\sigma }(i\nabla )-V_{p}(y).
\end{eqnarray*}
From (\ref{GH}) it follows that for \( \sigma =\sigma _{r,\delta } \)
\[
tr(Q_{p}(i\nabla ,y))_{-}\leq tr(Q_{p,\delta ,\sigma _{r,\delta }}(i\nabla ,y))_{-}\]
if \( r\leq \min \{\mu _{-},\mu _{+},C(\mu _{+},\mu _{-})\delta \} \). For
the eigenvalue sum on the right hand side we can apply (\ref{bdolm1}) in Lemma
\ref{sec: Lme1} and we conclude 

\begin{lem}
\label{sec: esabv}Assume that \( 0<r\leq \min \{\mu _{-},\mu _{+},C(\mu _{+},\mu _{-})\delta \} \)
and \( 0<\delta <1/2 \). Then the inequality
\begin{equation}
\label{esab}
S(p)\leq \int (Q_{p,\delta ,\sigma _{r,\delta }}(\xi ,y))_{-}d\gamma +\int _{0}^{\kappa }tr\, \chi _{-t}(Q_{p,\delta ,\sigma _{r,\delta }}(i\nabla ,y))dt
\end{equation}
holds true for some \( \kappa \leq c\sqrt{\vartheta (V)r^{-1}p^{-3}} \), where
the constant \( c \) in the estimate for \( \kappa  \) can be chosen to be
independent on \( V \), \( p \), \( M \), \( r \) and \( \delta  \).
\end{lem}

\section{The Proof of Theorem \ref{sec: maintm1}}

We are now in the position to complete the proof of formula (\ref{mainas1}).
In the beginning we shall assume that \( V \) has uniformly bounded second
derivatives and that \( V\in L^{\theta }(\mathbb R^{d})\cap L^{d+1}(\mathbb R^{d}) \)
for some \( \theta <\frac{d+1}{2} \) and \( d\geq 3 \).

\subsection{The estimate from above.}

First note that \( G_{p,\delta } \) is convex and consequently \( G_{p,\delta }(\xi )\leq G_{p,\delta ,\sigma _{r}}(\xi ) \)
for all \( \xi \in \mathbb {R}^{d} \). Thus,
\begin{eqnarray*}
\int (Q_{p,\delta ,\sigma _{r,\delta }}(\xi ,y))_{-}d\gamma  & \leq  & \int (G_{p,\delta }(\xi )-V_{p}(x))_{-}d\gamma \\
 & \leq  & (1-\delta )^{-1}\int (Q_{p}(\xi ,x))_{-}d\gamma =\frac{1}{1-\delta }\Sigma _{p}(V).
\end{eqnarray*}
Simultaneously we have
\begin{eqnarray*}
tr\, \chi _{-t}(Q_{p,\delta ,\sigma _{r,\delta }}(i\nabla ,y)) & \leq  & tr\, \chi _{-t}(G_{p,\delta }(i\nabla )-V_{p}(y))\\
 & \leq  & N_{p}((V((1-\delta )x_{1},x_{2},x_{3})-tp)_{+})
\end{eqnarray*}
for all \( t\geq 0 \). Hence, relations (\ref{esab}), (\ref{bd13}) and (\ref{bd13d4})
imply that
\[
(1-\delta )S_{p}(V)\leq \Sigma _{p}(V)+\co p(1+\ln \frac{p}{M})\int \min \left\{ V_{+}^{2},\kappa V_{+}\right\} dx+\frac{\co }{p}\left\Vert V_{+}\right\Vert _{L^{4}}^{4}\]
in the dimension \( d=3 \), or
\[
(1-\delta )S_{p}(V)\leq \Sigma _{p}(V)+\co p^{\frac{d-1}{2}}\int \min \left\{ V_{+}^{\frac{d+1}{2}},\kappa V_{+}^{\frac{d-1}{2}}\right\} dx+\frac{\co }{p}\left\Vert V_{+}\right\Vert _{L^{d+1}}^{d+1}\]
if \( d\geq 4 \), hold true for all \( p\geq M \) with \( \kappa =c\sqrt{\vartheta (V)r^{-1}p^{-3}} \).
Since \( V_{+}^{*}(t)\leq \co \left\Vert V_{+}\right\Vert _{L^{\theta }}t^{-\theta ^{-1}} \)
we find that
\begin{eqnarray*}
\int \min \left\{ V_{+}^{\frac{d+1}{2}},\kappa V_{+}^{\frac{d-1}{2}}\right\} dx & = & \int _{0}^{\infty }\min \left\{ (V_{+}^{*})^{\frac{d+1}{2}},\kappa (V_{+}^{*})^{\frac{d-1}{2}}\right\} dt\\
 & \leq  & \co \left\Vert V_{+}\right\Vert ^{\theta }_{L^{\theta }}\kappa ^{\frac{d+1}{2}-\theta },
\end{eqnarray*}
and consequently
\begin{equation}
\label{spo1}
S_{p}(V)\leq \frac{\Sigma _{p}(V)}{1-\delta }+\co \frac{p\left( 1+\ln \frac{p}{M}\right) \vartheta ^{\beta }(V)\left\Vert V_{+}\right\Vert ^{\theta }_{L^{\theta }}}{(1-\delta )p^{3\beta }r^{\beta }}+\co \frac{\left\Vert V_{+}\right\Vert _{L^{4}}^{4}}{(1-\delta )p}
\end{equation}
as \( 0<M\leq p \) with \( \beta =\frac{2-\theta }{2}>0 \) if \( d=3 \) and
\begin{equation}
\label{spo2}
S_{p}(V)\leq \frac{\Sigma _{p}(V)}{1-\delta }+\co \frac{p^{\frac{d-1}{2}}\vartheta ^{\beta }(V)\left\Vert V_{+}\right\Vert ^{\theta }_{L^{\theta }}}{(1-\delta )p^{3\beta }r^{\beta }}+\co \frac{\left\Vert V_{+}\right\Vert _{L^{d+1}}^{d+1}}{(1-\delta )p}
\end{equation}
as \( 0<M\leq p \) with \( \beta =\frac{d+1}{4}-\frac{\theta }{2}>0 \) if
\( d\geq 4 \). 

Pick now \( \delta (p)=p^{-\epsilon } \) and \( r=r(p)=\min \{\mu _{+},\mu _{-},C(\mu _{+},\mu _{-})\delta \} \)
with \( 0<\epsilon <3 \). Since \( \Sigma _{p}(V) \) is of order \( p^{\frac{d-1}{2}} \)
for large \( p \), we claim
\[
\limsup _{p\to \infty }p^{-1}S_{p}(V)\leq \lim _{p\to \infty }p^{-1}\Sigma _{p}(V).\]

\subsection{The estimate from below.}

On the other hand, from (\ref{st3}) and from the identity \( H_{p}(\xi )=H_{p,\sigma _{r}}(\xi ) \)
for \( \xi \in \mathbb R^{d}\backslash (B_{r}^{+}\cup B_{r}^{-}) \) it follows
that
\begin{eqnarray*}
S_{p}(V) & \geq  & \int (Q_{p,\sigma _{r}}+\kappa )_{-}d\gamma \\
 & \geq  & \Sigma _{p}(V-p\kappa )-\int _{y\in B^{+}_{r}\cup B_{r}^{-}}(Q_{p}+\kappa )_{-}d\gamma .
\end{eqnarray*}
Next note that at least \( \left[ \frac{\mu _{\pm }-r}{2r}\right]  \) disjoint
balls of radius \( r \) can be placed into the domains \( [r-\mu _{-},0]\times (-r,r)^{2} \)
and \( [0,\mu _{+}-r]\times (-r,r)^{2} \), respectively. Because of \( H_{p}(\eta ,\zeta )\geq H_{p}(\eta ^{\prime },\zeta ) \)
for all \( |\eta ^{\prime }|\leq |\eta | \) we can conclude that
\begin{eqnarray*}
\left[ \frac{\mu _{\pm }-r}{2r}\right] \int _{y\in B_{r}^{\pm }}(Q_{p}+\kappa )_{-}d\gamma  & \leq  & \int _{\xi \in [r-\mu _{-},1-\mu _{+}]\times (-r,r)^{2}}(Q_{p}+\kappa )_{-}d\gamma \\
 & \leq  & \Sigma _{p}(V-p\kappa )
\end{eqnarray*}
and
\begin{equation}
\label{spu}
S_{p}(V)\geq \left( 1-\frac{1}{\left[ \frac{\mu _{-}-r}{2r}\right] }-\frac{1}{\left[ \frac{\mu _{+}-r}{2r}\right] }\right) \Sigma _{p}(V-p\kappa ).
\end{equation}
Put now \( r=r(p)=p^{-\alpha } \) with \( 0<\alpha <1 \). Then \( r\to 0 \)
and simultaneously \( p\kappa =\co \vartheta ^{1/2}(V)r^{-1/2}p^{-1/2}\to 0 \)
as \( p\to \infty  \). Thus, it holds
\[
\Sigma _{p}(V-p\kappa )\geq \Sigma _{p}(V-\delta )\]
for arbitrary \( \delta >0 \) if \( p \) is large enough. Because of the given
class of potentials this means
\begin{eqnarray*}
\liminf _{p\to \infty }S_{p}(V) & \geq  & \lim _{p\to \infty }\Sigma _{p}(V-\delta )\\
 & \geq  & \frac{\omega _{d}}{(d+1)2^{\frac{3d-1}{2}}\pi ^{d}}\int _{\mathbb {R}^{d}}(V(x)-\delta )_{+}^{\frac{d+1}{2}}dx.
\end{eqnarray*}
Since \( V\in L^{\frac{d+1}{2}} \) we can pass to the limit \( \delta \to 0 \).

\subsection{The closure of the asymptotical formula.}

If \( d\geq 4 \), we finally apply inequality (\ref{bd245}) in a standard
manner to close asymptotics (\ref{mainas1}) to all potentials \( V_{+}\in L^{\frac{d+1}{2}}\cap L^{d+1} \).
However, for \( d=3 \) the appropriate Lieb-Thirring inequality (\ref{bd24})
contains the logarithmic factor \( 1+\ln \frac{p}{M} \), which prevents us
from carrying out the same procedure in that case.

\section{The Proof of Theorem \ref{sec: maintm2}}

For the proof of Theorem \ref{sec: maintm2} we follow the main strategy of
\cite{ELSS} and apply the bounds (\ref{spo1}), (\ref{spo2})and (\ref{spu})
of the previous section in a more subtle way. For the shortness of notation
we shall write
\[
Y_{p}=U(y;p)\chi _{0}(Q_{p}(i\nabla ,y))=pU_{p}\chi _{0}(Q_{p}(i\nabla ,y)),\]
where in agreement with our previous notation \( U_{p}(y)=p^{-1}U(yp^{-1}) \).

\subsection{The estimate from above.}

Let \( \{\psi _{p,n}\} \) be an o.n. system of eigenfunctions corresponding
to the negative part of \( Q_{p}(i\nabla ,y) \). Then for any \( \epsilon \in (0,1) \)
it holds
\begin{eqnarray*}
tr\: p^{-1}Y_{p} & = & \sum _{n}\int U_{p}(x)|\psi _{p,n}(x)|^{2}dx\\
 & \leq  & \frac{1}{\epsilon }\left( tr\: (Q_{p}(i\nabla ,y)-\epsilon U_{p})_{-}-tr\: (Q_{p}(i\nabla ,y))_{-}\right) .
\end{eqnarray*}
Here we make use of the variational property
\[
tr\: (Q_{p}(i\nabla ,y)-\epsilon U_{p})_{-}\geq tr\: D(\epsilon U_{p}-Q_{p}(i\nabla ,y))\]
for any operator \( 0\leq D\leq 1 \). Put \( V_{\epsilon }=V+\epsilon U \).
Then (\ref{spo1}) - (\ref{spu}) imply that

\[
tr\: p^{-1}Y_{p}\leq \frac{1}{\epsilon }\left( \Sigma _{p}(V_{\epsilon })-\Sigma _{p}(V-p\kappa )+R(p,\epsilon ,\mu _{\pm },\delta ,V,U)\right) ,\]
for all \( 0<M\leq p \) and \( \epsilon \in (0,1) \), where
\begin{eqnarray*}
R(p,\epsilon ,\mu _{\pm },\delta ,V,U) & = & \frac{\delta }{1-\delta }\Sigma _{p}(V_{\epsilon })+\co \frac{\left\Vert V_{\epsilon }\right\Vert _{L^{d+1}}^{d+1}}{(1-\delta )p}\\
 &  & +\co \frac{p^{\frac{d-1}{2}}z_{d}(p)\vartheta ^{\beta }(V_{\epsilon })\left\Vert V_{\epsilon }\right\Vert ^{\theta }_{L^{\theta }}}{(1-\delta )p^{3\beta }r^{\beta }}+\co r\Sigma _{p}(V-p\kappa )
\end{eqnarray*}
with \( \beta =\frac{d+1}{4}-\frac{\theta }{2} \), \( z_{3}(p)=1+\ln \frac{p}{M} \)
and \( z_{d}(p)=1 \) for \( d\geq 4 \). 

Pick now \( \delta (p)=p^{-\alpha } \) and \( r=r(p)=\min \{\mu _{-},\mu _{+},C(\mu _{-},\mu _{+})\delta (p)\} \)
with \( 0<\alpha <1 \). Fix \( \epsilon \in (0,1) \). Then \( r\to 0 \),
\( p\kappa \to 0 \) and the limits
\begin{eqnarray*}
p^{-\frac{d-1}{2}}\Sigma _{p}(V_{\epsilon }) & \to  & \frac{\omega _{d}}{(d+1)2^{\frac{3d-1}{2}}\pi ^{d}}\int V_{\epsilon }^{\frac{d+1}{2}}dx,\\
p^{-\frac{d-1}{2}}\Sigma _{p}(V) & \to  & \frac{\omega _{d}}{(d+1)2^{\frac{3d-1}{2}}\pi ^{d}}\int V^{\frac{d+1}{2}}dx,\\
p^{-\frac{d-1}{2}}\Sigma _{p}(V-p\kappa ) & \to  & \frac{\omega _{d}}{(d+1)2^{\frac{3d-1}{2}}\pi ^{d}}\int V^{\frac{d+1}{2}}dx
\end{eqnarray*}
hold true as \( p\to \infty  \). From this we conclude that
\[
\limsup _{p\to \infty }p^{-\frac{d+1}{2}}tr\: Y^{p}\leq \frac{\epsilon ^{-1}\omega _{d}}{(d+1)2^{\frac{3d-1}{2}}\pi ^{d}}\left( \int V_{\epsilon }^{\frac{d+1}{2}}dx-\int V^{\frac{d+1}{2}}dx\right) .\]
Note that for non-negative \( U,V\in L^{\frac{d+1}{2}} \) and all \( \epsilon \in (0,1) \)
we have
\[
\epsilon ^{-1}|V_{\epsilon }^{\frac{d+1}{2}}-V^{\frac{d+1}{2}}|\leq \frac{d+1}{2}U(V+U)^{\frac{d-1}{2}},\]
where the function on the r.h.s. is integrable. Hence, by Lebegues' majorization
theorem we can pass to the limit \( \epsilon \to +0 \) and find
\[
\limsup _{p\to \infty }p^{-\frac{d+1}{2}}Y_{p}\leq \frac{\omega _{d}}{2^{\frac{3d+1}{2}}\pi ^{d}}\int UV^{\frac{d-1}{2}}dx.\]

\subsection{The estimate from below.}

Reversely, it holds
\[
tr\: p^{-1}Y_{p}\geq \frac{1}{\epsilon }\left( tr\: (Q_{p}(i\nabla ,y))_{-}-tr\: (Q_{p}(i\nabla ,y)+\epsilon U_{p})_{-}\right) .\]
Let \( V_{-\epsilon }=V-\epsilon U \) with \( \epsilon \in (0,1) \). Then
\[
tr\: p^{-1}Y_{p}\geq \frac{1}{\epsilon }\left( \Sigma _{p}(V)-\Sigma _{p}(V_{-\epsilon }-p\kappa )+\tilde{R}(p,\epsilon ,\mu _{\pm },\delta ,V,U)\right) ,\]
for all \( 0<M\leq p \) and \( \epsilon \in (0,1) \), where
\begin{eqnarray*}
\tilde{R}(p,\epsilon ,\mu _{\pm },\delta ,V,U) & = & \frac{\delta }{1-\delta }\Sigma _{p}(V)+\co \frac{\left\Vert V\right\Vert _{L^{d+1}}^{d+1}}{(1-\delta )p}\\
 &  & +\co \frac{p^{\frac{d-1}{2}}z_{d}(p)\vartheta ^{\beta }(V)\left\Vert V\right\Vert ^{\theta }_{L^{\theta }}}{(1-\delta )p^{3\beta }r^{\beta }}+\co r\Sigma _{p}(V_{-\epsilon }-p\kappa )
\end{eqnarray*}
with \( \beta =\frac{d+1}{4}-\frac{\theta }{2} \), \( z_{3}(p)=1+\ln \frac{p}{M} \)
and \( z_{d}(p)=1 \) for \( d\geq 4 \). Passing to \( p\to \infty  \) as
above we obtain
\[
\liminf _{p\to \infty }p^{-\frac{d+1}{2}}tr\: Y^{p}\geq \frac{\epsilon ^{-1}\omega _{d}}{(d+1)2^{\frac{3d-1}{2}}\pi ^{d}}\left( \int V^{\frac{d+1}{2}}dx-\int (V_{-\epsilon })_{+}^{\frac{d+1}{2}}dx\right) \]
and for \( \epsilon \to +0 \) by a majorized convergence argument this turns
into
\[
\liminf _{p\to \infty }p^{-\frac{d+1}{2}}Y_{p}\geq \frac{\omega _{d}}{2^{\frac{3d+1}{2}}\pi ^{d}}\int UV^{\frac{d-1}{2}}dx.\]

\section{Appendix I: Phase Space Estimates for the Symbol \protect\( Q_{p}(\xi ,y)=H_{p}(\xi )-V_{p}(y)\protect \). }

\subsection{Preliminaries.}

Let \( V \) be a real function on \( {\Bbb R}^{d} \). Set \( V_{p}(y)=p^{-1}V(p^{-1}y) \)
and
\[
Q_{p}(\xi ,y)=H_{p}(\xi )-V_{p}(y),\]
where
\[
H_{p}(\xi )=T_{+}(\xi )+T_{-}(\xi )-\sqrt{1+M^{2}p^{-2}}\]
for
\[
T_{\pm }(\xi )=\sqrt{|(\eta \mp \mu _{\pm })^{2}+|\zeta |^{2}+\mu ^{2}_{\pm }M^{2}p^{-2},}\]
with \( \xi \in {\Bbb R}^{d} \), \( \xi =(\eta ,\zeta ) \) for \( \xi _{1}=\eta \in {\Bbb R} \)
and \( (\xi _{2},\dots ,\xi _{d})=\zeta \in {\Bbb R}^{d-1} \), \( M=m_{+}+m_{-} \),
\( \mu _{\pm }=m_{\pm }M^{-1}>0 \), \( p>0 \). Below we shall study properties
of the phase space averages
\begin{eqnarray}
\Sigma _{p}=\Sigma _{p}(V) & = & (2\pi )^{-d}\int \int (Q_{p}(\xi ,y))_{-}d\xi dy,\label{Sup} \\
\Xi _{p}=\Xi _{p}(V) & = & (2\pi )^{-d}\int \int _{Q_{p}<0}d\xi dy.\label{Nup} 
\end{eqnarray}
Set
\begin{equation}
\label{Ladef}
\Lambda _{p}(y;V)=(2\pi )^{-d}\int _{Q_{p<0}}d\xi .
\end{equation}

\begin{lem}
\label{sec: Lphv}Assume that \( \tau =Mp^{-1}\leq 1 \). Then for any \( y\in \Bbb R^{d} \)
it holds
\begin{equation}
\label{Lambda}
\Lambda _{p}(y)=\frac{\omega _{d}W^{\frac{d}{2}}\left( W+\upsilon \right) \left( W+2\upsilon \right) ^{\frac{d}{2}}\left( W^{2}+2W\upsilon +\tau ^{2}(1-4\tilde{\mu }^{2})\right) ^{\frac{d}{2}}}{(4\pi )^{d}\left( W^{2}+2W\upsilon +\tau ^{2}\right) ^{\frac{d+1}{2}}},
\end{equation}
where \( W=W(y)=(V_{p}(y))_{+} \) and \( \upsilon =\sqrt{1+\tau ^{2}} \).
\end{lem}
\begin{proof}
Fix some point \( y\in \Bbb R^{d} \). Since \( H_{p}(\xi )\geq 0 \) we have
\( Q_{p}(\xi ,y)\geq 0 \) if \( V_{p}(y)\leq 0 \), what settles the statement
in that case. Assume now \( V_{p}(y)\geq 0 \). Put \( \tilde{\mu }=(\mu _{-}-\mu _{+})/2 \)
and \( \tilde{\eta }=\eta +\tilde{\mu } \). Then \( Q_{p}(\xi ,y)<0 \) is
equivalent to
\begin{eqnarray}
\notag  &  & 2\sqrt{\left( |\zeta |^{2}+\tilde{\eta }^{2}+\frac{1}{4}+\left( \tilde{\mu }^{2}+\frac{1}{4}\right) \tau ^{2}\right) ^{2}-(\tilde{\eta }-\tilde{\mu }\tau ^{2})^{2}}\\
 & < & A^{2}-2\left( |\zeta |^{2}+\tilde{\eta }^{2}+\frac{1}{4}+\left( \tilde{\mu }^{2}+\frac{1}{4}\right) \tau ^{2}\right) ,\label{c} 
\end{eqnarray}
where \( A=V_{p}+\sqrt{1+\tau ^{2}} \). Thus, in particular, the condition
\begin{equation}
\label{co}
|\zeta |^{2}+\tilde{\eta }^{2}<B,\quad B=\frac{A^{2}}{2}-\left( \tilde{\mu }^{2}+\frac{1}{4}\right) \tau ^{2}-\frac{1}{4}
\end{equation}
has to be satisfied. The bound (\ref{c}) transforms into
\begin{equation}
\label{ellip}
|\zeta |^{2}+\frac{A^{2}-1}{A^{2}}\left( \tilde{\eta }+\frac{\tilde{\mu }\tau ^{2}}{A^{2}-1}\right) ^{2}<B-\frac{A^{2}}{4}+\frac{\tilde{\mu }^{2}\tau ^{4}}{A^{2}-1},
\end{equation}
subject to the additional condition (\ref{co}). For \( V_{p}(y)\geq 0 \) we
have \( A\geq \sqrt{1+\tau ^{2}} \) and inequality (\ref{ellip}) describes
an ellipsoid with symmetry semi-axes of the length
\begin{eqnarray*}
l_{1} & = & \frac{A}{\sqrt{A^{2}-1}}\sqrt{B-\frac{A^{2}}{4}+\frac{\tilde{\mu }^{2}\tau ^{4}}{A^{2}-1}}\\
l_{2}=\cdots =l_{d} & = & \sqrt{B-\frac{A^{2}}{4}+\frac{\tilde{\mu }^{2}\tau ^{4}}{A^{2}-1}}.
\end{eqnarray*}
It is not difficult to see that \( l_{j}^{2}\leq B \) for \( j=2,\dots ,d \)
and \( \tau \leq 1 \), while \( l_{1}\leq B^{1/2}-\frac{\tilde{\mu }\tau ^{2}}{A^{2}-1} \).
Thus the ellipsoid given by (\ref{ellip}) is a subset of the sphere (\ref{co}),
and the volume of all admissible \( \xi  \) is given by \( \omega _{d}l_{1}\dots l_{d} \),
what by
\[
B-\frac{A^{2}}{4}+\frac{\tilde{\mu }^{2}\tau ^{4}}{A^{2}-1}=\frac{(A^{2}-1-\tau ^{2})(A^{2}-1-4\tilde{\mu }^{2}\tau ^{2})}{4(A^{2}-1)}\]
implies the second statement of the Lemma \ref{sec: Lphv}.
\end{proof}
Let us now assume that \( \tau =Mp^{-1}\leq 1 \) and \( m_{\pm }>0 \). Then
\( \tilde{\mu }^{2}<1/4 \) and by (\ref{Lambda}) the quantity \( \Lambda _{p} \)
permitts the following two-sided estimate
\begin{equation}
\label{prelest}
\Lambda _{p}(y;V)\asymp \left\{ \begin{array}{lcl}
\tau ^{-1}(V_{p}(y))_{+}^{d/2} & \quad \mbox {on}\quad  & \Omega _{1}=\{y|V_{p}(y)\leq \tau ^{2}\}\\
(V_{p}(y))^{\frac{d-1}{2}}_{+} & \quad \mbox {on}\quad  & \Omega _{2}=\{y|\tau ^{2}\leq V_{p}(y)\leq 1\}\\
(V_{p}(y))_{+}^{d} & \quad \mbox {on}\quad  & \Omega _{3}=\{y|V_{p}(y)\geq 1\}
\end{array}\right. ,
\end{equation}
or equivalently,
\begin{equation}
\label{prel1}
\Lambda _{p}(y;V)\asymp \min \left\{ \tau ^{-1}(V_{p}(y))_{+}^{\frac{d}{2}},(V_{p}(y))^{\frac{d-1}{2}}_{+}\right\} +(V_{p}(y))_{+}^{d},
\end{equation}
which for fixed \( \tilde{\mu } \) is uniform for all \( p \) and \( M \)
satisfying \( \tau \leq 1 \). Hence, \( V_{+}\in L^{d/2}(\mathbb {R}^{d})\cap L^{d}(\mathbb {R}^{d}) \)
is sufficient and necessary for
\begin{equation}
\label{Ladef1}
\Xi _{p}(V)=\int \Lambda _{p}(y;V)dy=p^{d}\int \Lambda _{p}(px;V)dx
\end{equation}
to be finite.

\subsection{Potentials \protect\( V_{+}\in L^{\frac{d-1}{2}}(\mathbb {R}^{d})\cap L^{d}(\mathbb {R}^{d})\protect \)}

For this class of potentials by (\ref{prelest}) the function \( p^{\frac{d-1}{2}}\Lambda _{p}(p\cdot ) \)
has an integrable majorant, and by Lebesgues' limit theorem it holds
\begin{equation}
\label{dp1}
\lim _{p\to \infty }p^{-\frac{d+1}{2}}\Xi _{p}=\lim _{p\to \infty }\int p^{\frac{d-1}{2}}\Lambda _{p}(py)dy=\frac{\omega _{d}}{2^{\frac{3d+1}{2}}\pi ^{d}}\int (V_{+}(y))^{\frac{d-1}{2}}dy.
\end{equation}

\subsection{Potentials \protect\( V_{+}\in L^{\frac{d+1}{2}}(\mathbb {R}^{d})\cap L^{d+1}(\mathbb {R}^{d})\protect \)}

We find that the integrand on the r.h.s. of
\[
p^{\frac{1-d}{2}}\Sigma _{p}(V)=p^{\frac{1-d}{2}}\int _{0}^{\infty }\Xi _{p}(V-sp)ds=\int _{0}^{\infty }\int _{\mathbb {R}^{d}}p^{\frac{d-1}{2}}\Lambda _{p}(py;V-t)d^{3}ydt\]
is for fixed \( \tilde{\mu } \) bounded by a uniform multiple of
\[
\max \{(V(y)-t)^{\frac{d-1}{2}}_{+},(V(y)-t)_{+}^{d}\},\]
which is integrable on \( [0,\infty )\times \mathbb {R}^{d} \) for \( V_{+}\in L^{\frac{d+1}{2}}(\mathbb {R}^{d})\cap L^{d+1}(\mathbb {R}^{d}) \).
Thus,
\begin{eqnarray}
\lim _{p\to \infty }p^{-\frac{d-1}{2}}\Sigma _{p}(V) & = & \int _{0}^{\infty }dt\int _{\mathbb {R}^{d}}\lim _{p\to \infty }\left( p^{\frac{d-1}{2}}\Lambda _{p}(py;V-t)\right) dy\notag \\
 & = & \frac{\omega _{d}}{2^{\frac{1+3d}{2}}\pi ^{d}}\int _{0}^{\infty }dt\int _{\mathbb {R}^{d}}(V(y)-t)^{\frac{d-1}{2}}_{+}dy\notag \\
 & = & \frac{\omega _{d}}{(d+1)2^{\frac{3d-1}{2}}\pi ^{d}}\int _{\mathbb {R}^{d}}V_{+}^{\frac{d+1}{2}}(y)dy.\label{sp24} 
\end{eqnarray}

\subsection{Potentials \protect\( V_{+}\in L^{\theta }_{w}(\mathbb {R}^{d})\cap L^{d}(\mathbb {R}^{d})\protect \)
with \protect\( \frac{d-1}{2}<\theta <\frac{d}{2}\protect \)}

For potentials \( V \) where \( V_{+} \) is ``strictly between'' \( L^{\frac{d-1}{2}}(\mathbb {R}^{d})\cap L^{d}(\mathbb {R}^{d}) \)
and \( L^{\frac{d}{2}}(\mathbb {R}^{d})\cap L^{d}(\mathbb {R}^{d}) \) the phase
space volume shows a different behaviour in \( p \). Let us study the model
potential
\begin{equation}
\label{vy}
V(y)=\min \{1,v|y|^{-d/\theta }\},
\end{equation}
where \( \frac{d-1}{2}<\theta <\frac{d}{2} \). Then \( V=V_{+}\in L^{\theta }_{w}(\mathbb {R}^{d}) \)
and \( \left\Vert V\right\Vert _{\theta ,w}=c(\theta ,d)v \). The preliminary
estimate (\ref{prelest}) shows that
\begin{eqnarray*}
\Xi _{p} & \asymp  & p^{\frac{1-d}{2}}\int _{p^{1-\frac{\theta }{d}}v^{\frac{\theta }{d}}\leq |y|\leq pv^{\frac{\theta }{d}}}dy\\
 &  & +v^{\frac{d-1}{2}}p^{\frac{(d-1)(d-\theta )}{2\theta }}\int _{pv^{\frac{\theta }{d}}\leq |y|\leq p^{1+\frac{\theta }{d}}v^{\frac{\theta }{d}}M^{-\frac{2\theta }{d}}}|y|^{-\frac{d(d-1)}{2\theta }}dy\\
 &  & +v^{\frac{d}{2}}p^{1-\frac{d}{2}+\frac{d^{2}}{2\theta }}M^{-1}\int _{|y|\geq p^{1+\frac{\theta }{d}}v^{\frac{\theta }{d}}M^{-\frac{2\theta }{d}}}|y|^{-\frac{d^{2}}{2\theta }}dy\\
 & \asymp  & p^{\theta +1}v^{\theta }M^{d-1-2\theta }(1+o(1))
\end{eqnarray*}
as \( p\to \infty  \). After one has established the order of \( \Xi _{p} \)
in \( p \), the same estimate now shows that
\[
\Xi _{p}=(1+o(1))\int _{|y|>p^{1+c}}\Lambda _{p}(y)dy,\quad 0<c<\frac{\theta }{d},\]
as \( p\to \infty  \). Hence, for \( \frac{d-1}{2}<\theta <\frac{d}{2} \)
it holds
\begin{eqnarray*}
\Xi _{p} & = & (1+o(1))\int _{|y|>p^{1+c}}\Lambda _{p}(y)dy\\
 & = & (1+o(1))\frac{\omega _{d}}{(4\pi )^{d}}\int _{|y|>p^{1+c}}W^{\frac{d}{2}}\frac{2^{\frac{d}{2}}(2W+\tau ^{2}\hat{\mu })^{\frac{d}{2}}}{(2W+\tau ^{2})^{\frac{d+1}{2}}}dy,
\end{eqnarray*}
where \( \hat{\mu }=1-4\tilde{\mu }^{2}\in (0,1] \) and \( W(y)=p^{-1}\min \left\{ 1,vp^{\frac{d}{\theta }}|y|^{-\frac{d}{\theta }}\right\}  \)
as \( p\to \infty  \). This implies
\[
\Xi _{p}=(1+o(1))\frac{2^{\frac{d}{2}}\omega _{d}p^{d+1}}{(4\pi )^{d}M}\int _{|x|>p^{c}}\left( \frac{v}{p}\right) ^{\frac{d}{2}}|x|^{-\frac{d^{2}}{2\theta }}\frac{\left( 2\frac{vp}{M^{2}}|x|^{-\frac{d}{\theta }}+\hat{\mu }\right) ^{\frac{d}{2}}}{\left( 2\frac{vp}{M^{2}}|x|^{-\frac{d}{\theta }}+1\right) ^{\frac{d+1}{2}}}dx\]
or
\begin{equation}
\label{tpl1}
\Xi _{p}=(1+o(1))\frac{2^{\theta }\theta \omega _{d}^{2}}{(4\pi )^{d}}v^{\theta }M^{d-1-2\theta }L(d,\theta ,\hat{\mu })p^{\theta +1}\quad \mbox {as}\quad p\to \infty 
\end{equation}
with
\begin{eqnarray*}
L(d,\theta ,\hat{\mu }) & = & \int _{0}^{\infty }\frac{(t+\hat{\mu })^{\frac{d}{2}}t^{\frac{d}{2}-\theta -1}dt}{(t+1)^{\frac{d-1}{2}}}\\
 & = & \hat{\mu }^{d-\theta }B\left( \frac{d}{2}-\theta ,\theta -\frac{d-1}{2}\right) \! _{2}F_{1}\left( \frac{d}{2}-\theta ,\frac{d+1}{2};\frac{1}{2},1-\hat{\mu }\right) ,
\end{eqnarray*}
where \( _{2}F_{1} \) is Gauss´ hypergeometric function (\cite{PBM} 2.2.6.24
p.303).This result can be generalised to all potentials \( V \) with \( V_{+}\in L^{\theta }_{w}\cap L^{d} \)
for \( \frac{d-1}{2}<\theta <\frac{d}{2} \) and \( \Delta _{\theta }(V_{+})=\delta _{\theta }(V_{+})=c(\theta ,d)v \).

\subsection{Potentials \protect\( V_{+}\in L^{\theta }_{w}(\mathbb {R}^{d})\cap L^{d}(\mathbb {R}^{d})\protect \)
with \protect\( \frac{d+1}{2}<\theta <\frac{d+2}{2}\protect \)}

A similar calculation can be carried out for the average \( S_{p}(V) \) if
the potential (\ref{vy}) satisfies \( \frac{d+1}{2}<\theta <\frac{d}{2}+1 \)
and is therefore ``strictly'' between \( L^{\frac{d+1}{2}}\cap L^{d} \) and
\( L^{\frac{d}{2}+1}\cap L^{d} \). First, from (\ref{prelest}) one concludes
in general that
\begin{eqnarray}
\Sigma _{p} & = & \int _{0}^{\infty }\int _{\mathbb {R}^{d}}\Lambda _{p}(y;V-sp)dyds\label{Thl0} \\
 & \asymp  & \int _{\mathbb {R}^{d}}\Theta _{p}(y;V)dy,\label{Thl} 
\end{eqnarray}
for sufficient large \( p \), where
\[
\Theta _{p}(y;V)=(V_{p}(y))_{+}^{d+1}+\tau ^{d+1}\chi _{\Omega _{2}\cup \Omega _{3}}(y)+\tau ^{-1}(V_{p}(y))_{+}^{\frac{d}{2}+1}\chi _{\Omega _{1}}(y)\]
with \( \chi  \) being the characteristic functions of (unions of) the respective
sets \( \Omega _{j} \) defined in (\ref{prelest}). For the potential \( V(y)=\min \{1,v|y|^{-\frac{d}{\theta }}\} \)
at hand this gives the preliminary estimate
\[
\Sigma _{p}\asymp p^{\theta -1}M^{d+1-2\theta }v^{\theta }\quad \mbox {as}\quad p\to \infty .\]
Moreover, the integration in (\ref{Thl}) and therefore in (\ref{Thl0}) can
be reduced to \( |y|>p^{1+c} \), \( 0<c<\frac{\theta }{d} \), without changing
the asymptotical behaviour of the integrals. Hence, if \( \frac{d+1}{2}<\theta <\frac{d}{2}+1 \),
\( \phi =(1+o(1))2^{\frac{d}{2}}(4\pi )^{-d}\omega _{d} \) as \( p\to \infty  \)
and \( a=1+cd\theta ^{-1} \), we have
\begin{eqnarray*}
\Sigma _{p} & = & \phi \int _{0}^{\infty }\int _{|y|\geq p^{1+c}}\frac{(V_{p}-s)_{+}^{\frac{d}{2}}(2(V_{p}-s)_{+}+\tau ^{2}\hat{\mu })^{\frac{d}{2}}dyds}{(2(V_{p}-s)_{+}+\tau ^{2})^{\frac{d+1}{2}}}\\
 & = & \phi d\omega ^{d}p^{d}\int _{0}^{\infty }\int ^{\infty }_{p^{c}}\frac{(\frac{v}{p}r^{-\frac{d}{\theta }}-s)_{+}^{\frac{d}{2}}(2(\frac{v}{p}r^{-\frac{d}{\theta }}-s)_{+}+\tau ^{2}\hat{\mu })^{\frac{d}{2}}r^{d-1}drds}{(2(\frac{v}{p}r^{-\frac{d}{\theta }}-s)_{+}+\tau ^{2})^{\frac{d+1}{2}}}\\
 & = & \phi \theta \omega ^{d}v^{\theta }p^{d-\theta }\int _{0}^{\infty }\int ^{vp^{-a}}_{0}\frac{(t-s)_{+}^{\frac{d}{2}}(2(t-s)_{+}+\tau ^{2}\hat{\mu })^{\frac{d}{2}}t^{-\theta -1}dtds}{(2(t-s)_{+}+\tau ^{2})^{\frac{d+1}{2}}}.
\end{eqnarray*}
The later integral can be simplified as follows
\begin{eqnarray*}
\Sigma _{p} & = & \phi \theta \omega ^{d}v^{\theta }p^{d-\theta }\int _{0}^{vp^{-a}}dtt^{-\theta -1}\int _{0}^{t}\frac{x^{\frac{d}{2}}(2x+\tau ^{2}\hat{\mu })^{\frac{d}{2}}}{(2x+\tau ^{2})^{\frac{d+1}{2}}}dx\\
 & = & \phi 2^{-\frac{d+3}{2}}\theta \omega ^{d}v^{\theta }\tau ^{d+1}p^{d-\theta }\int _{0}^{vp^{-a}}dtt^{-\theta -1}\int _{0}^{2t\tau ^{-2}}\frac{u^{\frac{d}{2}}(u+\hat{\mu })^{\frac{d}{2}}}{(u+1)^{\frac{d+1}{2}}}du\\
 & = & \phi 2^{\theta -\frac{d+3}{2}}\theta \omega ^{d}v^{\theta }\tau ^{d+1-2\theta }p^{d-\theta }\times \\
 &  & \qquad \times \int _{0}^{2vM^{-2}p^{2-a}}dww^{-\theta -1}\int _{0}^{w}\frac{u^{\frac{d}{2}}(u+\hat{\mu })^{\frac{d}{2}}}{(u+1)^{\frac{d+1}{2}}}du.
\end{eqnarray*}
For \( a<2 \) and \( \theta >\frac{d+1}{2} \) we finally claim
\begin{equation}
\label{tmi1}
\Sigma _{p}=(1+o(1))\frac{2^{\theta -\frac{3}{2}}\theta \omega ^{2}_{d}}{(4\pi )^{d}}v^{\theta }M^{d+1-2\theta }p^{\theta -1}K(d,\theta ,\hat{\mu }),
\end{equation}
where \( K(d,\theta ,\hat{\mu }) \) denotes the finite positive constant
\[
K(d,\theta ,\hat{\mu })=\int _{0}^{\infty }dww^{-\theta -1}\int _{0}^{w}\frac{u^{\frac{d}{2}}(u+\hat{\mu })^{\frac{d}{2}}}{(u+1)^{\frac{d+1}{2}}}du.\]
In fact, this asymptotics holds true for all \( V_{+}\in L^{\theta }_{w}\cap L^{d} \),
\( \frac{d+1}{2}<\theta <\frac{d}{2}+1 \), with \( \Delta _{\theta }(V_{+})=\delta _{\theta }(V_{+})=c(\theta ,d)v \).

\section{Appendix II: An Estimate \protect\( N_{p}(V)\leq c(V)p^{2}\protect \) in the
Dimension \protect\( d=3\protect \).}

\subsection{Statement of the result.}

In this appendix we show, that for certain short-range potentials with some
repulsive tail at infinity the counting function \( N_{p}(V) \) in the dimension
\( d=3 \) is bounded by a multiple of \( p^{2} \). This complements the estimate
(\ref{esnp1}). As above we concentrate on the case of positive masses \( m_{\pm }>0 \).

\begin{thm}
\label{tm: A1}Assume that \( d=3 \), \( m_{\pm }>0 \) and that the bounded
potential \( V \) satisfies the condition
\begin{equation}
\label{A.1.1}
V(x)\leq -a(1+|x|-b)^{-\gamma },\quad x\in \mathbb R^{3},\: |x|\geq b,
\end{equation}
for appropriate positive finite constants \( a \), \( b \) and \( \gamma  \).
Then
\begin{equation}
\label{A.1.2}
N_{p}(V)\leq C(b+1)^{3}p^{2},\qquad p\geq M,
\end{equation}
where \( C=C(a,\gamma ,\left\Vert V\right\Vert _{L^{\infty }}) \) does not
depend on \( p \) and \( b \).
\end{thm}

\subsection{A localization estimate in spatial coordinates. }

Consider the operator
\[
T=\sqrt{-\Delta +1}\quad \mbox {on}\quad L^{2}(\mathbb R^{3}).\]
Let \( (\cdot ,\cdot ) \) and \( \left\Vert \cdot \right\Vert  \) be the scalar
product and the norm in \( L^{2}(\mathbb R^{3}) \). For positive \( b \) and
\( \gamma  \) set \( \varsigma _{\gamma ,b}(x)=(1+|x|-b)^{-\gamma /2} \),
\( x\in \mathbb R^{3} \). The proof of Theorem \ref{tm: A1} is based on the
following improved localization estimate:

\begin{lem}
\label{lm A2}For any given positive number \( b \) one can find spherically
symmetric functions \( \chi _{1},\chi _{2}\in C^{2}(\mathbb R^{3}) \), which
are monotone w.r.t. the radial variable and satisfy
\begin{equation}
\label{A.2}
\chi _{1}(x)=1\quad \mbox {if}\quad |x|\leq b,\quad \chi _{1}(x)=0\quad \mbox {if}\quad |x|\geq b+1,\quad \chi _{1}^{2}+\chi _{2}^{2}=1,
\end{equation}
such that for any \( \epsilon >0 \) and \( \gamma >0 \) the estimate 
\begin{equation}
\label{A.3}
\left| (Tu,u)-\sum _{j=1}^{2}(Tu\chi _{j},u\chi _{j})\right| \leq c\left\Vert u\chi _{1}\right\Vert ^{2}+\epsilon \left\Vert u\chi _{2}\varsigma _{\gamma ,b}\right\Vert ^{2}
\end{equation}
holds true for all \( u\in C_{0}^{\infty }(\mathbb R^{3}) \) with some appropriate
finite constant \( c=c(\gamma ,\epsilon ) \).
\end{lem}
\begin{proof}
For given \( b>0 \) we can obviously chose spherically symmetric cut-off functions
\( \chi _{1},\chi _{2}\in C^{2}(\mathbb R^{3}) \), which are monotone in the
radial variable and satisfy (\ref{A.2}) as well as
\begin{equation}
\label{A.4}
\chi _{1}(x)\chi _{2}(x)>0\quad \mbox {for}\quad b<|x|<b+1.
\end{equation}
According to formula (3.8) in \cite{LY} the localization error of the operator
\( T \) is given as follows
\begin{equation}
\label{A.5}
(Tu,u)-\sum _{j=1}^{2}(Tu\chi _{j},u\chi _{j})=(Lu,u),
\end{equation}
where \( L \) is an integral operator with the kernel
\[
L(x,y)=\frac{K_{2}(|x-y|)\sum _{j=1}^{2}(\chi _{j}(x)-\chi _{j}(y))^{2}}{(2\pi )^{2}|x-y|^{2}}.\]
Here \( K_{2} \) stands for the modified Bessel function and satisfies the
estimate
\begin{equation}
\label{A.6}
|K_{2}(|x-y|)|\leq \alpha |x-y|^{-2}e^{-\kappa |x-y|}
\end{equation}
for appropriate \( \alpha ,\kappa >0 \). 

We shall now estimate the quadratic form on the r.h.s. of (\ref{A.5}). Because
of symmetry it suffices to estimate the respective integrals over the region
\( |x|\leq |y| \) only. Let \( \delta \in (0,1/2) \) be a positive number,
which will be specified later. Put
\[
b_{\delta }=b+1-\delta \]
and define
\begin{eqnarray*}
O_{1} & = & \{(x,y)|\: |x|\leq |y|\leq b_{\delta }\},\\
O_{2} & = & \{(x,y)|\: |x|\leq b_{2\delta },\: |y|\geq b_{\delta }\},\\
O_{3} & = & \{(x,y)|\: |x|\leq |y|,\: |x|\geq b_{2\delta },\: (x,y)\not \in O_{1}\cup O_{2}\}.
\end{eqnarray*}
Then
\begin{equation}
\label{A.9}
(Lu,u)_{L^{2}(\mathbb R^{3})}=2\mbox {Re}(I_{1}+I_{2}+I_{3}),
\end{equation}
where
\[
I_{k}=\iint _{O_{k}}L(x,y)u(y)\bar{u}(x)dxdy,\quad k=1,2,3.\]

To estimate \( I_{1} \) we notice that
\begin{equation}
\label{A.10}
|\chi _{j}(x)-\chi _{j}(y)|\leq \min \left\{ 1,\co |x-y|\right\} \quad \mbox {for\: all}\quad x,y\in \mathbb R^{3}.
\end{equation}
From (\ref{A.6}) and (\ref{A.10}) we conclude
\[
|L(x,y)|\leq \co |x-y|^{-2}\min \{1,|x-y|^{-2}\}\quad \mbox {for\: all}\quad (x,y)\in O_{1}.\]
Hence, it holds
\begin{eqnarray*}
|I|_{1} & \leq  & 2^{-1}\iint _{(x,y)\in O_{1}}(|u(x)|^{2}+|u(y)|^{2})|L(x,y)|dxdy\\
 & \leq  & \co \int _{|x|\leq b_{\delta }}|u(x)|^{2}dx\int _{\mathbb R^{3}}|x-y|^{-2}\min \{1,|x-y|^{-2}\}dy.
\end{eqnarray*}
This gives
\begin{equation}
\label{A.12}
|I_{1}|\leq \co \int _{|x|\leq b_{\delta }}|u(x)|^{2}dx\leq \co (\delta )\left\Vert u\chi _{1}\right\Vert ^{2}_{L^{2}(\mathbb R^{3})}.
\end{equation}
In the last step we used that \( \chi _{1}(x)\geq \co (\delta )>0 \) for all
\( |x|\leq b_{\delta } \), what on its turn follows from (\ref{A.4}) and the
radial monotonicity of \( \chi _{1} \). 

We study now the integral \( I_{2} \) and observe that
\begin{equation}
\label{A.13}
|L(x,y)|\leq \co \delta ^{-2}e^{-\kappa |x-y|}\quad \mbox {for}\quad (x,y)\in O_{2}.
\end{equation}
In view of
\[
|u(y)u(x)|\leq 4^{-1}\epsilon _{1}^{-1}|u(x)|^{2}+\epsilon _{1}|u(y)|^{2},\quad \epsilon _{1}>0,\]
we find from (\ref{A.13}) that for any given \( \gamma >0 \) and \( \epsilon _{1}>0 \)
the bound
\begin{eqnarray}
|I_{2}| & \leq  & \co \delta ^{-2}\epsilon _{1}^{-1}\int _{|x|\leq b_{2\delta }}dx|u(x)|^{2}\int _{|y|\geq b_{\delta }}e^{-\kappa |x-y|}dy\notag \\
 &  & +\co \frac{\epsilon _{1}}{\delta ^{2}}\int _{|y|\geq b_{\delta }}\frac{dy|u(y)|^{2}}{e^{\frac{\kappa }{2}(|y|-b_{2\delta })}}\int _{|x-y|\geq \delta }e^{-\frac{\kappa }{2}|x-y|}dx\label{A.14.0} 
\end{eqnarray}
holds true. By (\ref{A.4}) and by the radial monotonicity of \( \chi _{1} \)
and \( \chi _{2} \) we have
\begin{eqnarray*}
\chi _{1}(x)\geq \co (\delta )>0 & \quad \mbox {for}\quad  & |x|\leq b_{2\delta },\\
\chi _{2}(y)\geq \co (\delta )>0 & \quad \mbox {for}\quad  & |y|\geq b_{\delta }>b.
\end{eqnarray*}
Moreover, it holds
\[
e^{-\frac{\kappa }{4}(|y|-b_{2\delta })}\leq \co (\gamma ,\delta )\varsigma _{\gamma ,b}(y),\quad |y|\geq b_{\delta }.\]
Hence, the inequality (\ref{A.14.0}) implies
\begin{equation}
\label{A.14}
|I_{2}|\leq \co (\delta ,\epsilon _{1})\left\Vert u\chi _{1}\right\Vert ^{2}_{L^{2}(\mathbb R^{3})}+\co (\gamma ,\delta )\epsilon _{1}\left\Vert u\chi _{2}\varsigma _{\gamma ,b}\right\Vert ^{2}_{L^{2}(\mathbb R^{3})}.
\end{equation}

Estimating \( I_{3} \) we recall that
\begin{equation}
\label{A.15.1}
\chi _{1}(x)\equiv 0\quad \mbox {and}\quad \chi _{2}(x)\equiv 1\quad \mbox {for\: all}\quad |x|\geq b+1.
\end{equation}
Since \( \chi _{1},\chi _{2}\in C^{2}(\mathbb R^{2}) \), for any given \( \epsilon _{2}>0 \)
we can find an appropriate \( \delta =\delta (\epsilon _{2})\in (0,1/2) \)
such that
\begin{equation}
\label{A.15.0}
|\nabla \chi _{1}(x)|^{2}+|\nabla \chi _{2}(x)|^{2}\leq \epsilon _{2},\quad b_{2\delta }\leq |x|\leq b+1.
\end{equation}
With this value of \( \delta  \) the relations (\ref{A.15.1}) and (\ref{A.15.0})
imply
\begin{equation}
\label{A.15.2}
\sum _{j=1}^{2}(\chi _{j}(x)-\chi _{j}(y))^{2}\leq \epsilon _{2}\min \left\{ 4\delta ^{2},|x-y|^{2}\right\} ,\quad b_{2\delta }\leq |x|\leq |y|.
\end{equation}
Moreover, from (\ref{A.13}), (\ref{A.15.1}) and (\ref{A.15.2}) we conclude
that
\[
|L(x,y)|\leq \epsilon _{2}\co |x-y|^{-2}\min \left\{ 4\delta ^{2}|x-y|^{-2},1\right\} \min \left\{ e^{-\kappa (|y|-b-1)},1\right\} \]
for \( b_{2\delta }\leq |x|\leq |y| \) and \( L(x,y)=0 \) for \( b+1\leq |x|\leq |y| \)
. Therefore it holds
\begin{eqnarray*}
|I_{3}| & \leq  & 2^{-1}\iint _{(x,y)\in O_{3}}(|u(x)|^{2}+|u(y)|^{2})|L(x,y)|dxdy\\
 & \leq  & \epsilon _{2}\co \int _{|x|\leq b+1}|u(x)|^{2}dx\int _{\mathbb R^{3}}|x-y|^{-2}\min \{4\delta ^{2}|x-y|^{-2},1\}dy\\
 &  & +\epsilon _{2}\co \int _{|y|\geq b_{2\delta }}\frac{|u(y)|^{2}dy}{e^{\kappa (|y|-b-1)}}\int _{\mathbb R^{3}}\frac{\min \{4\delta ^{2}|x-y|^{-2},1\}}{|x-y|^{2}}dx.
\end{eqnarray*}
Since \( e^{-\frac{\kappa }{2}(|y|-b-1)}\leq \co (\gamma ,\delta )\varsigma _{\gamma ,b}(y) \)
for \( |y|\geq b_{2\delta } \) and \( \delta \in (0,1/2) \), we conclude that
\begin{equation}
\label{A.16}
|I_{3}|\leq \epsilon _{2}\co (\gamma ,\delta )(\left\Vert u\chi _{1}\right\Vert _{L^{2}(\mathbb R^{3})}^{2}+\left\Vert u\chi _{2}\varsigma _{\gamma ,b}\right\Vert ^{2}_{L^{2}(\mathbb R^{3})}).
\end{equation}

We proceed now as follows. For given \( \epsilon >0 \) chose \( \epsilon _{2}>0 \)
such that the total constant in front of the bracket in (\ref{A.16}) for given
\( b \) and \( \gamma  \) does not exceed \( \epsilon /4 \). Fix the corresponding
\( \delta (\epsilon _{2})>0 \) for (\ref{A.15.0}) and subsequently (\ref{A.16})
to be satisfied. Finally, fix \( \epsilon _{1}>0 \) such that the total constant
in front of the term \( \left\Vert u\chi _{2}\varsigma _{\gamma ,b}\right\Vert _{L^{2}(\mathbb R^{3})}^{2} \)
in (\ref{A.14}) for given \( b \), \( \gamma  \) and \( \delta (\epsilon _{2}) \)
does not exceed \( \epsilon /4 \). Then (\ref{A.5}) together with (\ref{A.10}),
as well as (\ref{A.12}), (\ref{A.14}) and (\ref{A.16}) yield (\ref{A.3}).
\end{proof}
\begin{rem}
Let \( t_{rel}=t_{rel}(P) \) be the regularized kinetic part of the operator
(\ref{hrel}) on \( L^{2}(\mathbb R^{3}) \), that is
\begin{equation}
\label{trelop}
t_{rel}=\sqrt{|\mu _{+}P-i\nabla |^{2}+m_{+}^{2}}+\sqrt{|\mu _{-}P+i\nabla |^{2}+m_{-}^{2}}-\sqrt{p^{2}+M^{2},}
\end{equation}
where \( M>0 \), \( \mu _{\pm }=m_{\pm }M^{-1}>0 \), and \( P\in \mathbb R^{3} \),
\( p=|P| \). As an immediate consequence of (\ref{A.2}) in Lemma \ref{lm A2}
we find that for arbitrary positive \( \epsilon  \) and \( \gamma  \) it holds
\begin{equation}
\label{trel}
\left| (t_{rel}u,u)-\sum _{j=1}^{2}(t_{rel}u\chi _{j},u\chi _{j})\right| \leq c(\gamma ,\epsilon ,\mu _{j},M)\left\Vert u\chi _{1}\right\Vert ^{2}+\epsilon \left\Vert u\chi _{2}\varsigma _{\gamma ,b}\right\Vert ^{2}.
\end{equation}
The constant \( c(\gamma ,\epsilon ,\mu _{j},M) \) can be chosen to be independent
on \( P \) and \( b \).
\end{rem}

\subsection{A local estimate in momentum space.}

Let
\[
t_{rel}(\xi ,P)=\sqrt{|\mu _{+}P-\xi |^{2}+m_{+}^{2}}+\sqrt{|\mu _{-}P+\xi |^{2}+m_{+}^{2}}-\sqrt{p^{2}+M^{2}}\]
be the symbol of the operator (\ref{trelop}) where \( M>0 \), \( \mu _{\pm }=m_{\pm }M^{-1}>0 \)
and \( P,\xi \in \mathbb R^{3} \). Put \( \xi =(\eta ,\zeta ) \) with \( \eta \in \mathbb R \)
and \( \zeta \in \mathbb R^{2} \). We recall that \( P=(p,0,0) \) and \( \mu _{+}+\mu _{-}=1 \).

\begin{lem}
\label{lm3A}Assume that \( p\geq \nu \geq M \) and that \( \xi =(\eta ,\zeta ) \)
satisfies
\begin{equation}
\label{etazeta}
\xi \in W(\nu ,p)=\{\xi |(|\eta |\geq 3p)\}\cup \{\xi |(|\zeta |^{2}\geq \nu p)\}.
\end{equation}
Then
\begin{equation}
\label{trellm}
t_{rel}(\xi ,P)\geq 2^{-1}3^{-1/2}\nu .
\end{equation}

\end{lem}
\begin{proof}
Assume first \( |\eta |\geq 3p\geq 3\nu \geq 3M \). Then
\begin{equation}
\label{A.18}
t_{rel}(\xi ,P)\geq \sqrt{4p^{2}+M^{2}}-\sqrt{p^{2}+M^{2}}\geq 3(\sqrt{5}+\sqrt{2})^{-1}\nu .
\end{equation}
If instead \( |\zeta |^{2}\geq \nu p \) from \( \nu \geq M \) it follows that
\begin{eqnarray*}
t_{rel}(\xi ,P) & \geq  & \sqrt{p^{2}+|\zeta |^{2}+M^{2}}-\sqrt{p^{2}+M^{2}}\\
 & \geq  & 2^{-1}|\zeta |^{2}(p^{2}+|\zeta |^{2}+M^{2})^{-1/2}\\
 & \geq  & 2^{-1}\nu (1+\nu p^{-1}+\nu ^{2}p^{-2})^{-1/2}.
\end{eqnarray*}
Since \( p\geq \nu  \) we conclude \( t_{rel}(\xi ,P)\geq 2^{-1}3^{-1/2}\nu  \).
Together with (\ref{A.18}) this completes the proof.
\end{proof}

\subsection{The proof of Theorem \ref{tm: A1}.}

Let
\[
q_{rel}(P)=t_{rel}(P)-V(y),\quad P=(p,0,0),\]
be the operator (\ref{qrell}) for \( d=3 \). Obviously the total multiplicity
of the negative eigenvalues of this operator coincides with \( N_{p}(V) \).
To verify (\ref{A.1.2}) it suffices to construct a subspace \( G \) in \( L^{2}(\mathbb R^{3}) \)
of finite dimension \( \dim G\leq Cb^{3}p^{2} \) such that
\begin{equation}
\label{qrelpos}
(q_{rel}(P)u,u)_{L^{2}(\mathbb R^{3})}\geq 0\quad \mbox {for\: all}\quad u\in G_{0}^{\bot },
\end{equation}
where \( G_{0}^{\bot } \) is a \( q_{rel}(P) \)-form dense subset of \( G^{\bot }=L^{2}(\mathbb R^{3})\ominus G \).

For given \( b \) construct the cut-off functions \( \chi _{1} \), \( \chi _{2} \)
from Lemma \ref{lm A2}. Set \( n=(n_{1},n_{2},n_{3}) \) for \( n_{j}\in \mathbb N_{+} \)
and \( x=(x_{1},x_{2},x_{3}) \) with \( x_{j}\in \mathbb R \), \( j=1,2,3 \).
Let \( b^{\prime }=b+1 \). We define
\[
u_{n}(x)=\left\{ \begin{array}{ccc}
b^{\prime -\frac{3}{2}}\prod _{j=1}^{3}\sin \pi n_{j}\left( \frac{1}{2}+\frac{x_{j}}{b^{\prime }}\right)  & \: \mbox {for}\:  & |x_{j}|\leq b^{\prime },\: j=1,2,3,\\
0 &  & \mbox {otherwise}.
\end{array}\right. \]
Let \( \tilde{G}=\tilde{G}(\tau _{1},\tau _{\bot }) \) be the linear span of
all \( u_{n} \) where
\[
n_{1}\leq \tau _{1}b^{\prime }p,\qquad n_{2,3}\leq \tau _{\bot }b^{\prime }p^{1/2},\]
and the positive real numbers \( \tau _{1} \), \( \tau _{\bot } \) will be
specified below. We put
\[
G=G(\tau _{1},\tau _{\bot })=\{u|u=\tilde{u}\chi _{1},\: \tilde{u}\in \tilde{G}\}.\]
Obviously we have
\[
\dim G=\dim \tilde{G}\leq \tau _{1}\tau ^{2}_{\bot }b^{\prime 3}p^{2}.\]

To verify (\ref{qrelpos}) we first notice, that from the boundedness of \( V \),
(\ref{A.1.1}) and (\ref{trel}) it follows that
\begin{equation}
\label{qpos}
(q_{rel}(P)u,u)_{L^{2}(\mathbb R^{3})}\geq (t_{rel}(P)u\chi _{1},u\chi _{1})_{L^{2}(\mathbb R^{3})}-\tilde{c}\left\Vert u\chi _{1}\right\Vert _{L^{2}(\mathbb R^{3})}^{2}
\end{equation}
for all \( u\in C_{0}^{\infty }(\mathbb R^{3}) \). For fixed \( \chi _{1} \)
the constant \( \tilde{c}=\tilde{c}(a,\gamma ,\left\Vert V\right\Vert _{L^{\infty }}) \)
does not depend on \( p \), \( b^{\prime } \) or \( u \). Let \( W(\nu ,p) \)
be the set defined in (\ref{etazeta}) of Lemma \ref{lm3A} for the choice \( \nu =2^{3}3^{1/2}\tilde{c} \).
Below we shall show, that for appropriate constants \( \tau _{1}=\tau _{1}(\tilde{c}) \)
and \( \tau _{\bot }=\tau _{\bot }(\tilde{c}) \), which do not depend on \( p \),
the bound
\begin{equation}
\label{toprove}
\left\Vert \widehat{{u\chi _{1}}}\right\Vert _{W(\nu ,p)}\geq 2^{-1}\left\Vert \widehat{{u\chi _{1}}}\right\Vert _{L^{2}(\mathbb R^{3})},\quad u\bot G(\tau _{1},\tau _{\bot }),
\end{equation}
holds true. From (\ref{toprove}) and (\ref{trellm}) we conclude that
\begin{eqnarray*}
(t_{rel}(P)u\chi _{1},u\chi _{1})_{L^{2}(\mathbb R^{3})} & \geq  & (t_{rel}(\xi ,P)\widehat{{u\chi _{1}}}(\xi ),\widehat{{u\chi _{1}}}(\xi ))_{L^{2}(W(\nu ,p))}\\
 & \geq  & 4\tilde{c}\left\Vert \widehat{{u\chi _{1}}}\right\Vert ^{2}_{L^{2}(W(\nu ,p))}\\
 & \geq  & \tilde{c}\left\Vert u\chi _{1}\right\Vert _{L^{2}(\mathbb R^{3})}^{2},
\end{eqnarray*}
where \( u\bot G(\tau _{1},\tau _{\bot }) \) and \( u\in C_{0}^{\infty }(\mathbb R^{3}) \).
Together with (\ref{qrelpos}) and (\ref{qpos}) the later bound settles the
proof.

In the remaining part of this section we establish (\ref{toprove}). Consider
some function \( u\bot G \). Then \( u\chi _{1}\bot \tilde{G} \) and consequently
\( u\chi _{1}=\sum _{j=1}^{3}\sigma _{j} \), where \( \sigma _{j}=\sum _{n\in \Upsilon _{j}}c_{n}u_{n} \)
and
\begin{eqnarray*}
\Upsilon _{1} & = & \{n|n_{1}\geq \tau _{1}b^{\prime }p\},\\
\Upsilon _{2} & = & \{n|(n_{1}<\tau _{1}b^{\prime }p)\}\cap \{n|(n_{2}\geq \tau _{\bot }b^{\prime }p^{1/2})\},\\
\Upsilon _{3} & = & \{n|(n_{1}<\tau _{1}b^{\prime }p)\}\cap \{n|(n_{2}<\tau _{\bot }b^{\prime }p^{1/2})\}\cap \{n|(n_{3}\geq \tau _{\bot }b^{\prime }p^{1/2})\}.
\end{eqnarray*}
Put \( \tilde{W}=\mathbb R^{3}\backslash W(\nu ,p) \). Since
\[
\left\Vert \widehat{{u\chi _{1}}}\right\Vert _{L^{2}(\tilde{W})}\leq \sum _{j=1}^{3}\left\Vert \hat{\sigma }_{j}\right\Vert _{L^{2}(\tilde{W})},\]
for (\ref{toprove}) it suffices to show that
\begin{equation}
\label{7.4}
\left\Vert \hat{\sigma }_{j}\right\Vert _{L^{2}(\tilde{W})}\leq 6^{-1}\left\Vert \widehat{{u\chi _{1}}}\right\Vert _{L^{2}(\mathbb R^{3})},\quad j=1,2,3.
\end{equation}

We shall verify (\ref{7.4}) for \( j=1 \). The proof for the cases \( j=2,3 \)
is similar. Obviously we have \( \hat{\sigma }_{1}=\sum _{n\in \Upsilon _{1}}c_{n}\hat{u}_{n} \)
where
\[
\hat{u}_{1}(\xi )=b^{\prime -\frac{3}{2}}\prod _{j=1}^{3}e^{i\pi (n_{j}+\frac{1}{2})}\frac{\pi n_{j}}{b^{\prime }}\frac{\sin (\xi _{j}b^{\prime }-\frac{\pi n_{j}}{2})}{\xi _{j}^{2}-\frac{\pi ^{2}n^{2}_{j}}{4b^{\prime 2}}}.\]
Since
\[
\left\Vert \hat{\sigma }_{j}\right\Vert ^{2}_{L^{2}(\tilde{W})}\leq \int _{|\xi _{1}|<3p}|\hat{\sigma }_{1}(\xi )|^{2}d\xi ,\]
after integration in \( \zeta =(\xi _{2},\xi _{3}) \) and using the notation
\( \eta =\xi _{1} \), we find that
\begin{equation}
\label{si1}
\left\Vert \hat{\sigma }_{j}\right\Vert ^{2}_{L^{2}(\tilde{W})}\leq \frac{\pi ^{2}}{b^{\prime 3}}\int _{|\eta |<3p}\sum _{\begin{array}{c}
n_{2},n_{3}\in \mathbb N_{+}\\
n_{1},n_{1}^{\prime }\geq \tau _{1}b^{\prime }p
\end{array}}\frac{|c_{(n_{1},n_{2},n_{3})}c_{(n^{\prime }_{1},n_{2},n_{3})}|n_{1}n^{\prime }_{1}d\eta }{\left| \eta ^{2}-\frac{\pi ^{2}n_{1}^{2}}{4b^{\prime 2}}\right| \left| \eta ^{2}-\frac{\pi ^{2}n_{1}^{\prime 2}}{4b^{\prime 2}}\right| }.
\end{equation}
Let us assume that \( \tau _{1}^{2}\geq 72\pi ^{-2} \). Then we have \( \frac{\pi ^{2}n_{1}^{2}}{8b^{\prime 2}}\geq 9p^{2}\geq \eta ^{2} \)
and \( \frac{\pi ^{2}n_{1}^{\prime 2}}{8b^{\prime 2}}\geq 9p^{2}\geq \eta ^{2} \)
in the denominator in the previous sum and thus
\begin{equation}
\label{si2}
\int _{|\eta |\leq 3p}\frac{d\eta }{\left| \eta ^{2}-\frac{\pi ^{2}n_{1}^{2}}{4b^{\prime 2}}\right| \left| \eta ^{2}-\frac{\pi ^{2}n_{1}^{\prime 2}}{4b^{\prime 2}}\right| }\leq \frac{8b^{\prime 2}}{\pi ^{2}n_{1}^{2}}\cdot \frac{8b^{\prime 2}}{\pi ^{2}n^{\prime 2}_{1}}\cdot 6p.
\end{equation}
Applying Schwarz inequality in the summations over \( n_{1} \) and \( n_{1}^{\prime } \)
together with (\ref{si2}) to (\ref{si1}) we obtain
\[
\left\Vert \hat{\sigma }_{j}\right\Vert ^{2}_{L^{2}(\tilde{W})}\leq \frac{384b^{\prime }p}{\pi ^{2}}\left( \sum _{n_{1}\geq \tau _{1}b^{\prime }p}n_{1}^{-2}\right) \sum _{n\in \mathbb N_{+}^{3}}|c_{n}|^{2}.\]
Since \( \sum _{n_{1}\geq \tau _{1}bp}n_{1}^{-2}\leq 2(\tau _{1}b^{\prime }p)^{-1} \)
and \( \sum _{n\in \mathbb N_{+}^{3}}|c_{n}|^{2}=\left\Vert \widehat{{u\chi _{1}}}\right\Vert _{L^{2}(\mathbb R^{3})}^{2} \)
a choice of \( \tau _{1}=2\cdot 36\cdot 384\cdot \pi ^{-2}\geq 72\cdot \pi ^{-2} \)
will yield (\ref{7.4}) for \( j=1 \).

\email{\textsf{\tiny email: wugalter@rz.mathematik.uni-muenchen.de, Timo.Weidl@mathematik.uni-stuttgart.de}\tiny }

\end{document}